\documentclass[prd,preprintnumbers,
twocolumn,
eqsecnum,letterpaper,superscriptaddress,nofootinbib]{revtex4-1}
\usepackage[usenames,dvipsnames]{color}
\usepackage{amsmath,amssymb,graphicx}
\usepackage{tensor}
\usepackage{bm}
\usepackage{float}
\usepackage{times}
\usepackage{microtype}
\usepackage{booktabs}
\usepackage[normalem]{ulem}
\usepackage{todonotes}
\usepackage{listings}
\usepackage[varg]{txfonts}
\usepackage[colorlinks, pdfborder={0 0 0}]{hyperref}
\definecolor{LinkColor}{rgb}{0.75 , 0, 0}
\definecolor{CiteColor}{rgb}{0, 0.5, 0.5}
\definecolor{UrlColor}{rgb}{0, 0, 0.75}
\hypersetup{linkcolor=LinkColor}
\hypersetup{citecolor=CiteColor}
\hypersetup{urlcolor=UrlColor}
\allowdisplaybreaks
\usepackage[CaptionAfterwards]{fltpage}

\usepackage[section]{placeins}
\usepackage{multirow}

\usepackage{array}
\newcolumntype{L}[1]{>{\raggedright\let\newline\\\arraybackslash\hspace{0pt}}m{#1}}
\newcolumntype{C}[1]{>{\centering\let\newline\\\arraybackslash\hspace{0pt}}m{#1}}
\newcolumntype{R}[1]{>{\raggedleft\let\newline\\\arraybackslash\hspace{0pt}}m{#1}}

\definecolor{dkgreen}{rgb}{0,0.6,0}
\definecolor{gray}{rgb}{0.5,0.5,0.5}
\definecolor{mauve}{rgb}{0.58,0,0.82}

\begin{document}
\newcommand{\blue}{\color{blue}}
\newcommand{\red}{\color{red}}
\newcommand{\msun}{$M_{\odot}$}
\newcommand{\Msun}{M_{\odot}}
\newcommand{\vek}[1]{\boldsymbol{#1}}
\newcommand{\IGC}{Institute for Gravitation and the Cosmos, Department of Physics, Pennsylvania State University, University Park, PA 16802, USA}
\newcommand{\PSUPhys}{Department of Physics, The Pennsylvania State University, University Park, PA 16802, United States of America}
\newcommand{\TPI}{Theoretisch-Physikalisches Institut, Friedrich-Schiller-Universit\"at Jena, 07743, Jena, Germany}
\newcommand{\I}{\mathrm{i}}
\newcommand{\E}{\mathrm{e}}
\newcommand{\eff}{\text{eff}}
\newcommand{\Tc}{\, \text{,}}
\newcommand{\Td}{\, \text{.}}
\newcommand{\md}{\mathrm{d}}
\newcommand{\gwb}{\textsc{gwbench}}
\newcommand{\gwbS}{\textsc{gwbench }}
\newcommand{\network}{\texttt{Network}}
\newcommand{\networkS}{\texttt{Network} }
\newcommand{\detector}{\texttt{Detector}}
\newcommand{\detectorS}{\texttt{Detector} }
\newcommand{\waveform}{\texttt{Waveform} }
\newcommand{\lpf}{F_{lp}}
\newcommand{\ssb}{\textcolor{orange}}
\newcommand{\SSB}[1]{{\textcolor{orange}{\it\textbf{[SSB: #1]}} }}

\newcommand{\numberthis}{\stepcounter{equation}\tag{\theequation}}

\title{\gwb: a novel Fisher information package for gravitational-wave benchmarking}

\author{S.~Borhanian}
\affiliation{\IGC}
\affiliation{\TPI}
\email{ssohrab.borhanian@uni-jena.de}

\date{\today}

\begin{abstract} We present a new \textsc{Python} package, \gwb, 
implementing the well-established Fisher information formalism as a fast and
straightforward tool for the purpose of gravitational-wave benchmarking, i.e.
the estimation of signal-to-noise ratios and measurement errors of
gravitational waves observed by a network of detectors. Such an
infrastructure is necessary due to the high computational cost of Bayesian
parameter estimation methods which renders them less effective for the
scientific assessment of gravitational waveforms, detectors, and networks of
detectors, especially when determining their effects on large populations of
gravitational-wave sources spread throughout the universe. \gwbS further gives
quick access to detector locations and sensitivities, while including the
effects of Earth's rotation on the latter, as well as waveform models and their
derivatives, while giving access to the host of waveforms available in the LSC
Algorithm Library. With the provided functionality, \gwbS is relevant for a
wide variety of applications in gravitational-wave astronomy such as waveform
modeling, detector development, cosmology, and tests of general
relativity.\end{abstract}

\maketitle

\section{Introduction and motivation}\label{sec:intro}

The initial detections of gravitational waves (GWs) emitted during the
coalescences of black holes and neutron stars \cite{GW150914,GW170817} have
confirmed a fundamental prediction of general relativity and opened a new
observational window into the universe in the same breath. Further detections
by the LIGO and Virgo observatories have pushed the number of observed GW
signals into the decades \cite{GWTC1}, and hence set off the era of GW
astronomy. This success has ignited the desire for detections that allow
for better estimation of the information buried in the signals that the GW
observatories record, thus allowing us to reveal new phenomena and novel
physics as well as shed light into greater depths of the cosmos.

Such desires pose challenges on many essential components in the figurative GW
detection machinery, with the development of new waveform (WF) models and the
planning of future detector designs being two important tasks to tackle. The
former is imperative to lower the systematic biases in what kind of signals the
observatories will be able to detect, to increase the information that can be
extracted from the detector output, and possibly to lower the computational
cost of WF model evaluation. The inclusion of higher-order spherical harmonic
modes beyond the quadrupole \cite{London:2017bcn,Cotesta:2018fcv} is one such
example illustrating the impact of improved WF models as it enhances the
detectability of binaries whose orbits are highly inclined with respect to the
line of sight and allows for an improved parameter estimation by lifting the
degeneracy between luminosity distance and inclination angle.

The detector challenge is even more fundamental: new, ground-breaking analyses
require `richer' data to forge through. The current, so called
second-generation (2G) of GW detectors, LIGO Hanford and Livingston
\cite{aLIGO} as well as Virgo \cite{Virgo}, will be joined by two more
detectors, KAGRA \cite{KAGRA} and LIGO-India \cite{ligo_india}, and further
sensitivity improvements are planned to go in effect in the years to follow
\cite{aPlusWP, Aasi:2013wya}. Nevertheless the limits of these facilities to address future,
scientific goals are already apparent \cite{LISA_science_case,ET_science_case}.
Thus the GW community has been pursuing the conceptualization of the next
generation of detectors. The proposals range from the Voyager effort,
envisioning a multitude of upgrades to the LIGO facilities \cite{Voyager}, to
completely new, third-generation (3G) observatories, namely the Einstein
Telescope \cite{ET_1, ET_2, ET_science_case} and the Cosmic Explorer effort
\cite{CE_proposal}. In addition, efforts are under way to search for GW from
space, with the Laser Interferometer Space Antenna (LISA)
\cite{LISA_science_case}.

Independently of what WF models shall be developed and adopted, or which
facilities continue or start operating, the GW community will need tools to
\emph{benchmark} the scientific output of any combination of these two components
in a quick and efficient way. One such tool is the Fisher information formalism
(FIF) \cite{Cutler:1994ys,Poisson:1995ef} which has proven to be a viable
method for GW benchmarking. Its major caveat is its dependence on high
signal-to-noise ratios to give reliable results, thus making it mostly suitable
to exceptional events in current detectors.  Nevertheless, the formalism is
very useful and widely applicable in studies entailing future generations of
detectors. For an extended review on the caveats of the FIF we refer to
\cite{Vallisneri:2007ev}.
Further, developments to improve the speed of full Bayesian parameter estimation,
for example via the application of machine learning techniques
\cite{Chua:2019wwt, Gabbard:2019rde, Green:2020dnx, Green:2020hst}, could
make the Bayesian framework a viable benchmarking tool in the future.

In this work, we present a new \textsc{Python} package, \gwb, that implements
the FIF in an easy-to-use manner and further provides means to compute and
access a host of quantities that are necessary for benchmarking. For a selection
of GW detectors, a so-called \emph{network}, and given a WF model, \gwbS can
compute: plus and cross polarizations of the WF, detector power spectral
densities, antenna patterns, location phase factors, detector responses,
detector and network signal to noise ratios, and measurement errors in the WF
model parameters from the FIF. Two particularly substantial features are the
inclusion of WFs from the LSC Algorithm Library (\textsc{LAL}) \cite{lalsuite,wette2020}
and the capability to include the effects of Earth's rotation in the detector
antenna patterns.

The paper is structured as follows: Section~\ref{sec:trade_study}
introduces the Cosmic Explorer trade study which stimulated the development of
this package. In Section~\ref{sec:concepts} we briefly review the FIF and
aforementioned core concepts utilized by \gwb. Next, we show a straightforward
example application of \gwbS to showcase its functionality in
Section~\ref{sec:gwb_benchmarking} and proceed to present a few validation tests
of the core methods of \gwbS in Section~\ref{sec:validation}. In
Section~\ref{sec:applications}, we describe the application of the code in two
previous papers and conclude this work in Section~\ref{sec:summary}.


\section{Cosmic Explorer Trade Study}\label{sec:trade_study}

The Cosmic Explorer Project\footnote{https://cosmicexplorer.org/} aims to
develop among other things the science case for a next-generation GW detector,
the Cosmic Explorer (CE). CE is a promising US proposal for a ground-based
detector design which is targeting to answer a wide variety of scientific
questions \cite{Sathyaprakash:2019yqt, Kalogera:2019sui, Sathyaprakash:2019rom,
Sathyaprakash:2019nnu, Barack:2018yly} as part of a global network of GW
observatories. These questions are grouped under three overarching
\emph{science goals}: (i) dynamics of dense matter and extreme environments,
(ii) black holes and neutron stars through cosmic time, and (iii) extreme
gravity and fundamental physics. The questions encompass a broad range of
fields relevant to GW astronomy including but not restricted to neutron star
physics, cosmology, tests of general relativity, and the feasibility of
multi-messenger observations. 

\gwbS was developed to perform an exhaustive \emph{trade study} to gauge CE's
performance in reaching the set science goals. The trade study examines some 457
global GW detector networks combining 50 CE configurations with plausible
background detectors such as upgraded 2G facilities or the Einstein Telescope.
Besides the number of networks, the computational requirements are also driven
by the list of science questions to be answered.

Hence, we decided to implement the FIF in \gwb, as it allows us to perform the
required tasks effectively for many events and different waveform models in each
network of interest. The measurement error estimates computed from the FIF are
processed to obtain pre-defined performance metrics which in return allow us to
evaluate the science potential of the various network configurations.  Such
metrics are for example the signal-to-noise ratios, 90\%-credible sky areas,
expected event rates, and measurement error estimates on WF parameters, but also
on inferred quantities such as neutron star radii.

As such, \gwbS has an important role to play for the development of CE's science
case. The broad range of science goals that can be examined with the code
showcases that \gwbS is a powerful tool that has a wide variety of applications
in gravitational-wave astronomy. We present two of our recent studies that were
enabled by \gwbS in Section~\ref{sec:applications}.


\section{Concepts for gravitational wave benchmarking}\label{sec:concepts}

\gwbS implements a variety of standard concepts and aims to make them easily
accessible to the user. We briefly review them \cite{LRR_Sathyaprakash_2009} in
this section.

\paragraph*{Waveform polarizations:}

Gravitational waves are emitted by a host of astrophysical and cosmological
phenomena, with the coalescence of compact binaries as the most prominent
example. The incoming GWs can be decomposed into two polarizations, \emph{plus}
$h_+(t;\bm{\Theta}_{GW})$ and \emph{cross} $h_\times(t;\bm{\Theta}_{GW})$, and
carry an abundance of information about their source systems, encoded in the
parameter vector $\bm{\Theta}_{GW}$. This vector depends on the WF model. In
case of the coalescence of two black holes the standard set of such parameters,
$\bm{\Theta}_{GW} = \{m_1, m_2, \bm{\chi_1}, \bm{\chi_2}, D_L, \iota, t_c,
\phi_c\}$ contains the companion masses $m_1, m_2$ and dimensionless spin
vectors $\bm{\chi_1}, \bm{\chi_2}$, the binary's luminosity distance $D_L$,
the inclination angle $\iota$ of its orbital plane with respect to the line
of sight, as well as two integration parameters: the time $t_c$ and phase
$\phi_c$ of coalescence.

In the remainder of this manuscript, we will use the Fourier transforms of the
WF and other quantities in the frequency-domain:
\begin{align}
    H_{+/\times}(f) = \int_{-\infty}^{\infty}\md t\, h_{+/\times}(t)\,\E^{\I\,2\pi ft}.
\end{align}

\paragraph*{Antenna patterns and location phase factor:}

The quadrupolar nature of GWs results in sky-position dependent responses from
detectors to incoming waves. The responses are governed by the so-called
\emph{antenna pattern functions} $F_+$ and $F_\times$ which are functions of
the source's sky position---right ascension $\alpha$ and declination
$\delta$---and polarization angle $\psi$. Besides, the relative motion between
the source and detector results in a frequency-dependence in the
Fourier-domain \cite{Wen:2010cr,Zhao:2017cbb}; albeit being negligible for signals that are short compared to
the time scale of the motion.

Another group of important quantities that encode the spatial distribution of
detectors in networks are the \emph{location phase factors} $F_{lp}$ which
shift the phase of the detector responses appropriately for each detector:
\begin{align}
    F_{lp} = \exp\left[\I 2\pi f/c\, \hat{\bm{r}}(\alpha,\delta) \cdot \bm{d}(\alpha_d,\beta_d)\right],
\end{align}

\noindent where $\hat{\bm{r}}$ and $\bm{d}$ are the unit vector pointing to the
source and the position vector of the detector, respectively, if measured from the center
of Earth. $\alpha_d$ and $\beta_d$ are the detector's longitude and latitude,
respectively.

\paragraph*{Detector response:}

The aforementioned \emph{detector response} $H$ is the GW strain in the
frequency-domain measured by a detector, combining the WF polarizations with
the detector's antenna patterns and location phase factor. It differs from the
actually recorded detector output $s(t) = h(t) + n(t)$ in the time-domain or
$S(f) = H(f) + N(f)$ in the frequency-domain, which also contains contributions
from the detector noise $n(t)$ or $N(f)$, respectively. The detector response
is given by
\begin{eqnarray}
H(f;\bm{\Theta}) &=& F_{lp}(f;\alpha,\delta) \cdot \nonumber \\
&& [H_+(f;\bm{\Theta}_{GW})\,F_+(f;\alpha,\delta,\psi) + \nonumber  \\
&& \phantom{[}H_\times(f;\bm{\Theta}_{GW})\,F_\times(f;\alpha,\delta,\psi)],
\end{eqnarray}

\noindent where $\bm{\Theta} = \{\bm{\Theta}_{GW},\alpha,\delta,\psi\}$.

\paragraph*{Power spectral density:}

The sensitivity of a detector determines its ability to measure the strain $H$.
It is governed by systematics due to detector noise $n(t)$ and characterized by
the noise auto-correlation $\kappa = \overline{ n(t_1)\,n(t_2) }$\footnote{The
average is taken with respect to an ensemble of noise realizations. The times
$t_1$ and $t_2$ are any two arbitrary times at which the detector noise is
evaluated or measured.}. In the following, we assume the noise to be stationary
and Gaussian, with mean $\overline{n} = 0$. The former assumption ensures that
$\kappa$ only depends on the time-difference $t^* = t_1 - t_2$. Thus, the
\emph{one-sided power spectral density} (PSD) $S_n$---the Fourier transform of
$\kappa$ for positive frequencies and thus its frequency-domain
counterpart---is given by
\begin{eqnarray}
S_n(f) = \frac{1}{2}\int_{-\infty}^{\infty} \md t^* \kappa(t^*)\,\E^{\I 2\pi ft^*}, \text{  with }f>0.
\end{eqnarray}

\noindent $S_n(f)$ indicates the sensitivity of the detector at different GW frequencies.

\paragraph*{Noise-weighted scalar product and signal-to-noise ratio:}

The existence of noise in the detector output affects the detection and
measurement of the GW signal strain, obscuring the difference between various
detector responses to some level. One way to assess the \emph{overlap} of these
responses in a manner that takes the detector's sensitivity into account is by
defining a \emph{noise-weighted scalar product} \cite{LRR_Sathyaprakash_2009},
\begin{eqnarray}
\langle H,G\rangle = 2\int_0^{\infty}\md f \, \frac{H(f)G^*(f)+H^*(f)G(f)}{S_n(f)},
\end{eqnarray}

\noindent where $H,G$ are two detector responses, e.g. for two different WF
models or the same model evaluated at different parameters.

Given this scalar product, we can calculate the \emph{signal-to-noise ratio}
(SNR) $\rho$ of a given detector response $H$ for a given detector PSD $S_n$ as
\begin{eqnarray}
\rho^2 = \langle H,H \rangle = 4\int_0^{\infty}\md f \, \frac{|H(f)|^2}{S_n(f)},
\end{eqnarray}

\noindent indicating the relative loudness of the detector response compared to
the noise in the detector output \cite{LRR_Sathyaprakash_2009}.

\paragraph*{Fisher information formalism:}

The \emph{FIF} or \emph{Fisher analysis} is well-established in GW data
analysis and discussed in detail in \cite{Cutler:1994ys,Poisson:1995ef}. Here,
we give a brief summary of how it allows us to estimate parameter error bounds.

Since we assumed that the noise $n = s - h$ behaves Gaussian with zero
mean, the same holds for its Fourier transform $N = S - H$. Thus, the
probability of the noise can be written as
\begin{align}
p(\bm{\Theta}) = p^0(\bm{\Theta})\,\E^{-\frac{1}{2}\langle
S-H(\bm{\Theta}),S-H(\bm{\Theta})\rangle}, \label{eq:probability}
\end{align}

\noindent where $\bm{\Theta}$ and $p^0$ are the parameter vector of the
detector response $H$ and the prior on these parameters, respectively. Then the
value of $\bm{\Theta}$ at peak probability is a good estimate of the true value
$\bm{\Theta^*}$ for the given GW signal $S,$ assuming a large SNR.

The probability is the greatest when the exponential is the
largest\footnote{\cite{Poisson:1995ef} shows that this holds even for non-flat
priors $p^0$.}. Around the maximal value, the exponent $E = \langle S-H,
S-H\rangle$ can be expanded as
\begin{align}
E(\bm{\Theta}) = E(\bm{\Theta^*}) + \frac{1}{2} \left.\frac{\partial^2 E(\bm{\Theta})}{\partial \Theta_i \partial
\Theta_j}\right|_{\bm{\Theta}=\bm{\Theta^*}} \Delta\Theta_i\,\Delta\Theta_j + \dots,
\label{eq:exponent}
\end{align}

\noindent where $\Delta\Theta_i=\Theta_i-\Theta^*_i$. The Hessian
\begin{equation}
\frac{\partial^2 E(\bm{\Theta})}{\partial \Theta_i \partial\Theta_j} = 2 \,\langle
\partial_{\Theta_i}H(\bm{\Theta}), \partial_{\Theta_j}H(\bm{\Theta}) \rangle + \langle
\partial_{\Theta_i} \partial_{\Theta_j} H(\bm{\Theta}), N \rangle
\end{equation}

\noindent reduces for noise with zero-mean to the first summand in which we
identify the \emph{Fisher information matrix} (FIM) $\Gamma$
\begin{equation}
\Gamma_{ij}\equiv \langle \partial_{\Theta_i}H(\bm{\Theta}),
\partial_{\Theta_j}H(\bm{\Theta})\rangle . \label{eq:fisher}
\end{equation}

\noindent Equations \eqref{eq:exponent} and \eqref{eq:fisher} inserted in the
probability \eqref{eq:probability} yield
\begin{equation}
p(\bm{\Theta}) \sim \E^{-\frac{1}{2} \Gamma_{ij}\,\Delta \Theta_i \,\Delta\Theta_j}.
\end{equation}

\noindent Thus the assumption of Gaussianity motivates the association of the
FIM to be the inverse of the \emph{covariance matrix} $\Sigma \equiv
\Gamma^{-1}$. The diagonal and off-diagonal elements of $\Sigma$ denote the
variances and covariances of the parameters, respectively, due to the
uncertainty introduced by the detector noise and give $1\sigma$-error estimates
via $\sigma_{\Theta_i} = \sqrt{\Sigma_{ii}}$.

\paragraph*{Detector networks and associated quantities:}

The above definitions of the SNR $\rho$ and FIM $\Gamma$ depend on quantities
such as the detector response and PSD and are therefore inherently detector-specific.
In a network, detectors work in tandem to improve detectability and parameter
estimation which can be captured by defining \emph{network versions} of the SNR
and FIM
\begin{eqnarray}
\rho_\text{net}^2 &= \sum_{n=1}^{N_d}\,\rho_n^2, \\
\Gamma_\text{net} &= \sum_{n=1}^{N_d}\,\Gamma_n,
\end{eqnarray}

\noindent where $N_d$ is the number of detectors in the network and $\rho_n$
and $\Gamma_n$ are the SNR and FIM of the n-th detector, respectively.
Consequently, the network covariance matrix and error estimates follow from
$\Gamma_\text{net}$ as defined before.

\paragraph*{Condition number of the Fisher information matrix:}

If the covariance matrix is obtained via numerical inversion, the FIM needs to
be well-conditioned which can be checked with the condition number
$c_\Gamma = e_{max}/e_{min}$. Here, $e_{max},e_{min}$ are the largest and
smallest eigenvalues of $\Gamma$. If the condition number exceeds a critical
value---a conservative choice for double-precision is $c_{crit}=10^{15}$
\cite{Rodriguez:2011aa}---the inversion result cannot be trusted generically
and should thus be disregarded.


\section{Gravitational wave benchmarking with \gwb} \label{sec:gwb_benchmarking}

\subsection{Package overview}

The installation instructions for \gwbS and the source code are publicly
available on \textsc{GitLab}: \url{https://gitlab.com/sborhanian/gwbench}. The
package consists of 33 modules which implement the concepts discussed in
Section \ref{sec:concepts} as well as several WFs models.  In this section, we
will demonstrate with the aid of a few examples how to use the main module of
\gwb, \textsc{network}, to perform GW benchmarking. These examples are
available as \textsc{Python} scripts at
\url{https://gitlab.com/sborhanian/gwbench/-/tree/master/example_scripts}. 

\textsc{network} contains the \networkS class, see Appendix
\ref{app:net_vs_det}, which is designed to handle the concepts introduced in
Section \ref{sec:concepts}---detector initialization (antenna patterns,
location phase factors, and PSDs), WF evaluation, and both SNR and error
benchmarking---by conducting and facilitating the tasks of the other modules in
the package. We will refer to the appendices as needed to discuss these modules
and aspects of the package.

\subsection{Basic application using \gwb}\label{sec:gwb_application}

We begin with a step-by-step showcase of the basic usage of the \networkS
module following the example script \textsc{quick\_start.py}. In this example
we estimate the SNR, measurement errors, and 90\%-credible sky area for a signal
with the same masses as the first detected GW event GW150914 \cite{GW150914},
if it was measured by three detectors at the LIGO sites in Hanford, WA and
Livingston, LA and the Virgo site in Cascina, Italy. Each detector is set to
the aLIGO detector sensitivities \cite{aLIGO} and we perform the Fisher
analysis with the WF model TaylorF2 \cite{1991PhRvD..44.3819S, Poisson:1997ha,
Mikoczi:2005dn, Arun:2008kb, Buonanno:2009zt, Bohe:2013cla, Bohe:2015ana,
Mishra:2016whh}, see Appendix \ref{app:wf_models}.

We begin by importing the \textsc{numpy} and \textsc{gwbench.network} modules
and specify the \texttt{network\_spec} which tells the \networkS which
\detectorS instances to initialize: \texttt{['aLIGO\_H','aLIGO\_L','aLIGO\_V']}
will result in three detectors with aLIGO technology, located and oriented like
the Hanford, Livingston, and Virgo detectors (refer to the Appendices
\ref{app:locs} and \ref{app:tecs} for other sites and sensitivities available
in \gwb).

\begin{lstlisting}
import numpy as np
from gwbench import network

network_spec = ['aLIGO_H','aLIGO_L','aLIGO_V']
net = network.Network(network_spec)
\end{lstlisting}

\noindent While the \networkS class is the main interaction point for the user,
single-detector tasks are relayed to the \detectorS class. Appendix
\ref{app:net_vs_det} compares these two classes to show their similarities and
how they synergize to conduct the GW benchmarking together.

Next, we choose the WF model of interest, TaylorF2, and initialize the
\waveform object inside our \network.  \waveform is another important class
that handles WF selection and evaluation.  Appendix \ref{app:wf_models}
contains a brief description of this class and the full list of available WF
models in \gwb, at the time of writing this paper.

\begin{lstlisting}
wf_model_name = 'tf2'
net.set_wf_vars(wf_model_name=wf_model_name)
\end{lstlisting}

Finally, we need to set a few \networkS variables like the frequency array and
injection parameters in the detector frame, used for WF evaluation. Further,
the code needs to know with respect to which of these parameters to take
derivatives for the FIM.  Besides, we can set the \networkS to convert the
derivatives $\partial_p$ and errors $\sigma_p$ for certain parameters $p$ into
the cosine ($\partial_{\cos(p)}$ and $\sigma_{\cos(p)}$) or logarithmic
($\partial_{\log(p)}$ and $\sigma_{\log(p)}$) versions. The final option tells
the code whether to take Earth's rotation into account for the computation of
the antenna pattern functions, see Appendix \ref{app:locs}.

\begin{lstlisting}
f = np.arange(5.,61.5,2**-4)

inj_params = {
    'Mc':    30.9,
    'eta':   0.247,
    'chi1z': 0,
    'chi2z': 0,
    'DL':    475,
    'tc':    0,
    'phic':  0,
    'iota':  np.pi/4,
    'ra':    np.pi/4,
    'dec':   np.pi/4,
    'psi':   np.pi/4,
    'gmst0': 0
    }

deriv_symbs_string = 'Mc eta DL tc phic iota ra dec psi'

conv_cos = ('iota','dec')
conv_log = ('Mc','DL')

use_rot = 0

net.set_net_vars(
    f=f, inj_params=inj_params,
    deriv_symbs_string=deriv_symbs_string,
    conv_cos=conv_cos, conv_log=conv_log,
    use_rot=use_rot
    )
\end{lstlisting}

\noindent The high frequency cutoff of $61.5 \,\text{Hz}$ corresponds to the GW
frequency at the innermost stable circular orbit of a compact binary coalescence
with total detector-frame mass $M=\mathcal{M}/\eta^{3/5}=71.5\,M_\odot$ for the
example injections with detector-frame chirp mass $\mathcal{M}=30.9\,M_\odot$
and symmetric mass ratio $\eta=0.247$. The detector-frame masses correspond to
redshifted sources-frame masses of GW150914, i.e. $m_{\rm det} = m_{\rm source}
(1+z)$ where we used a fiducial value of $z=0.1$ and computed the luminosity
distance via Planck18 cosmology from the \textsc{astropy} package
\cite{astropy:2013,astropy:2018}. \texttt{gmst0} is the Greenwich Mean Sidereal
Time when the signal passes Earth's center.

The \networkS is prepared for benchmarking and we calculate the WF
polarizations and their derivatives with respect to the chosen parameters with
the following function calls.

\begin{lstlisting}
net.calc_wf_polarizations()
net.calc_wf_polarizations_derivs_num()
\end{lstlisting}

\noindent The first line only computes the plus and cross polarizations in the
frequency-domain, \texttt{net.hfp} and \texttt{net.hfc}, while the second
obtains a dictionary \texttt{net.del\_hfpc} containing the derivatives of the
two WF polarizations, in addition to that. The structure of these
\emph{derivative dictionaries} is detailed in Appendix \ref{app:deriv_dicts}.

Next, we setup antenna patterns, location phase factors, and PSDs for each
\detectorS in the \networkS.

\begin{lstlisting}
net.setup_ant_pat_lpf_psds()
\end{lstlisting}

\noindent The calculated quantities can be accessed in each \detectorS instance
via \texttt{net.detectors[i].Fp},  \texttt{net.detectors[i].Fc},
\texttt{net.detectors[i].Flp}, and \texttt{net.detectors[i].psd} for
$\texttt{i}=0,...,N_d-1$. We note that the \detectorS automatically truncates
the frequency array, if necessary, to the range dictated by the given detector's
PSD. This truncation is saved in \texttt{net.detectors[i].f} for
$\texttt{i}=0,...,N_d-1$ and translates to all consecutive computations which
depend on the frequency, e.g. in the detector response or the location phase
factor, and is necessary to avoid array-length mismatches within a \detectorS
instance. On the other hand, it can result in a mismatch when comparing
frequency-dependent quantities of two \detectorS instances or between the
\networkS and one of its \texttt{Detectors}.

With antenna patterns and location phase factors prepared, we can calculate the
detector responses and their derivatives, analogously to the case for the WF
polarizations.

\begin{lstlisting}
net.calc_det_responses()
net.calc_det_responses_derivs_num()
\end{lstlisting}

\noindent The evaluated responses and derivative dictionaries can be accessed
in each \detectorS instance via \texttt{net.detectors[i].hf} and
\texttt{net.detectors[i].del\_hf} for $\texttt{i}=0,...,N_d-1$.

With the previously computed PSDs, we obtain readily the detector and network SNRs.

\begin{lstlisting}
net.calc_snrs()
\end{lstlisting}

\noindent The function calculates both $\rho$ and $\rho^2$ for the total
network and each detector in the network. The values are saved in the
respective \networkS and \detectorS instances.

Finally, we compute the error estimates on the parameters specified in
\texttt{deriv\_symbs\_string} and the estimates of the 90\%-credible sky area.

\begin{lstlisting}
net.calc_errors()
net.calc_sky_area_90()

net.print_detectors()
net.print_network()
\end{lstlisting}

\noindent \texttt{calc\_errors} contains several steps: It starts by
calculating the FIMs for all the \detectorS instances and then the \networkS
one from those. Next, it computes and saves the condition number $c_\Gamma$ for
each FIM $\Gamma$, and checks if the matrices are well-conditioned
($c_\Gamma<10^{15}$). If this criterion is fulfilled, the code will proceed to
invert each FIM to obtain all the covariance matrices $\Sigma$ and consequently
the error estimates. Finally, it calculates the inversion accuracy $\epsilon =
||\Gamma\cdot\Sigma - I||_\text{\rm max}$, where $I$ and $||\cdot||_\text{\rm
max}$ are the identity and maximum matrix norm, respectively.

Analogously to the SNR computations, the \networkS and \detectorS instances
save their respective quantities. The final two function calls simply print the
contents of all the \detectorS instances inside the \networkS as well its own
contents.

\paragraph*{Further remarks:}

For the sake of simplicity, we omitted non-default function arguments in the
example above. The \networkS class is designed to perform
`default'-benchmarking after setting all the necessary variables via
\texttt{net.set\_wf\_vars} and \texttt{net.set\_net\_vars} without the need to
pass further variables during the consecutive function calls. Nevertheless,
certain functions give the user the option to tune the functionality of the
code. In the following, we will present these options for the main GW
benchmarking functions showcased in the example.

\paragraph*{\texttt{net.setup\_ant\_pat\_lpf\_psds}:}

This function takes the following three arguments (default values)\texttt{
F\_lo (-np.inf)},\texttt{ F\_hi (np.inf)}, and\texttt{ psd\_file\_dict (None)}.
The first two are user-settable frequency cutoffs which truncate the frequency
array, if they lie within the PSD's intrinsic frequency range.  The third
argument allows the user to specify PSD files that are not part of \gwb. It
has to be passed as

\begin{lstlisting}
psd_file_dict = {
    det_key1:{'psd_file':file1,'is_asd':bool1}, ...
    }
\end{lstlisting}

\noindent where the \texttt{det\_key} are the exact keys that each \detectorS
contains, the \texttt{file} are the relative paths to the PSD files, and the
booleans \texttt{bool} tell the code whether the file contains PSDs $S_n$
(\texttt{bool=0}) or amplitude spectral densities $\sqrt{S_n}$
(\texttt{bool=1}). The PSD files should be formatted like the txt-files
in \texttt{'gwbench/noise\_curves'}. A caveat of using external PSD files is
that the \networkS and \detectorS labels might not reflect this choice and will
need manual changes. Fortunately, these can be done quite easily, if needed, in
run time.

\gwbS can naturally be used to just compute the antenna patterns, location phase
factors, and PSDs of detectors as shown in the example script
\textsc{compute\_ant\_pat\_lpf\_psd.py}.

\paragraph*{\texttt{net.calc\_det\_responses\_derivs\_num, net.calc\_wf\_polarizations\_derivs\_num}:}

These functions calculate the derivatives of the detector responses and WF
polarizations numerically via the \textsc{numdifftools} package
\cite{numdifftools}. The numeric derivatives require four special arguments:
the step size \texttt{step}, the differentiation scheme \texttt{method}, the
differentiation order \texttt{order}, and the derivative order \texttt{n}. The
respective default values in \gwbS are \texttt{1e-7, 'central', 2,} and
\texttt{1}. We refer to the documentation of \textsc{numdifftools} for further
information. \gwbS further allows the use of symbolic derivatives for certain
waveforms via the \textsc{sympy} package. Their evaluation is faster, but
requires a few extra steps compared to the numeric ones. Their usage is
showcased in Section \ref{sec:sym}, while Section \ref{sec:num_vs_sym} compares
the output of the numeric and symbolic differentiation methods.

\paragraph*{\texttt{net.calc\_snrs, net.calc\_errors,} and \texttt{net.calc\_sky\_area\_90}:}

These three functions have a common argument \texttt{only\_net} which tells the
code whether to calculate the respective quantities just for the \networkS or
also for each \detectorS within. The default is \texttt{0}, computing both
network and detector quantities.

Besides, \texttt{net.calc\_errors} takes two more arguments (default values)
\texttt{cond\_sup (1e15)} and \texttt{by\_element (0)}. The first determines
the condition number supremum $c^*$ that the code uses to assess whether a
computed FIM $\Gamma$ is well-conditioned ($c_\Gamma < c^*$).  Passing
\texttt{None} sets it to infinity, effectively making every FIM
well-conditioned. The second argument tells the code whether to obtain a single
inversion error $\epsilon$ for the entire FIM via the maximum norm (default) or
$n$ errors $\epsilon_i$. $n$ is the dimension of the FIM while each
$\epsilon_i$ is the maximum of the elements of the $i$-th row and column
combined. The single inversion error $\epsilon$ is equal to the maximum of the
$n$ errors.

\subsection{Symbolic derivatives}\label{sec:sym}

\paragraph*{Compute dictionaries of lambdified \textsc{sympy} derivatives:}

While numeric derivatives are computed and evaluated in the same,
computationally expensive step, symbolic derivatives via \textsc{sympy} are
first computed generically and then evaluated for a given set of arguments. The
latter is faster than the numeric approach presented in Section
\ref{sec:gwb_application}, but requires the generation of the so-called
\emph{lambdified} derivative functions ahead of time. In the following, we will
showcase an example script that can handle this step:
\textsc{generate\_lambdified\_functions.py}.

The user has to specify the WF, the derivative variables, the detector locations of
interest, and whether to take Earth's rotation into account. By default, the
script is setup to (i) use the WF model TaylorF2, (ii) take derivatives with
respect to chirpmass $\mathcal{M}$, symmetric mass ratio $\eta$, luminosity
distance $D_L$, time and phase of coalescence $t_c,\, \phi_c$, inclination angle
$\iota$, right ascension $\alpha$, declination $\delta$, and polarization angle
$\psi$, (iii) use all locations available in \gwb, and (iv) include Earth's
rotation.

Next, the code initializes a \texttt{Waveform} object and checks whether the
derivative variables actually form a subset of the WF and antenna pattern
arguments. The actual computation is then handled by

\begin{lstlisting}
dr.generate_det_responses_sym(
    wf,deriv_symbs_string,locs=locs,use_rot=use_rot)
\end{lstlisting}

\noindent which takes the \texttt{Waveform}, the derivative variables, the
locations of interest, and the rotation setting.

The function generates two sets of dictionaries containing different
\emph{generic derivative expressions} stored as lambdified \textsc{sympy}
functions. The first type contains the derivatives of the WF polarizations
while the second stores the derivatives of the detector responses for the
detector locations specified in the script. These dictionaries are then saved in a
folder named \textsc{lambdified\_functions} using the \textsc{dill} package.
The output files contain one dictionary each and the file names are structured
as \\

\begin{lstlisting}
'par_deriv_WFM_{wf_model_name}_
    VAR_{variables}_DET_{location}.dat'
\end{lstlisting}

\noindent where \texttt{\{wf\_model\_name\}} and \texttt{\{variables\}} are set
according to the values passed to \texttt{generate\_det\_responses\_sym}.
\texttt{\{location\}} is replaced by \texttt{pl\_cr} in case of the WF
polarizations.

\paragraph*{Symbolic derivative evaluation for Fisher analysis:}

\textsc{sym\_gw\_benchmarking.py} showcases how to load and evaluate the
symbolic derivatives generated with \textsc{generate\_lambdified\_functions.py}
and how to use the \textsc{injections} module to generate random samples of
input parameters for the detector response evaluation (see Appendix
\ref{app:injections}). The script is mimicked by
\textsc{num\_gw\_benchmarking.py} which differs by computing the derivatives
numerically; it can be seen as a more generic expansion of the example shown in
Section \ref{sec:gwb_application}.

Most of the script simply prepares the benchmarking and hence we focus on the
usage of symbolic derivatives, in order to avoid repetition. As before, the
\networkS is the central hub and requires the same setup as shown in Section
\ref{sec:gwb_application} to prepare the \networkS and \detectorS instances. In
contrast to the earlier numeric example, the function call

\begin{lstlisting}
 net.calc_det_responses_derivs_num()
\end{lstlisting}

\noindent is replaced by

\begin{lstlisting}
 net.load_det_responses_derivs_sym()
 net.calc_det_responses_derivs_sym()
\end{lstlisting}

\noindent which first loads the generated lambdified derivative functions and
then evaluates them. This is all that is needed besides ensuring that the
variables \texttt{wf\_model\_name}, \texttt{wf\_other\_var\_dic},
\texttt{deriv\_symbs\_string}, and \texttt{use\_rot} are set to the same values
as during the generation of the lambdified derivatives.


\section{Validating \gwb}\label{sec:validation}

\begin{figure*}
    \includegraphics[width=\textwidth]{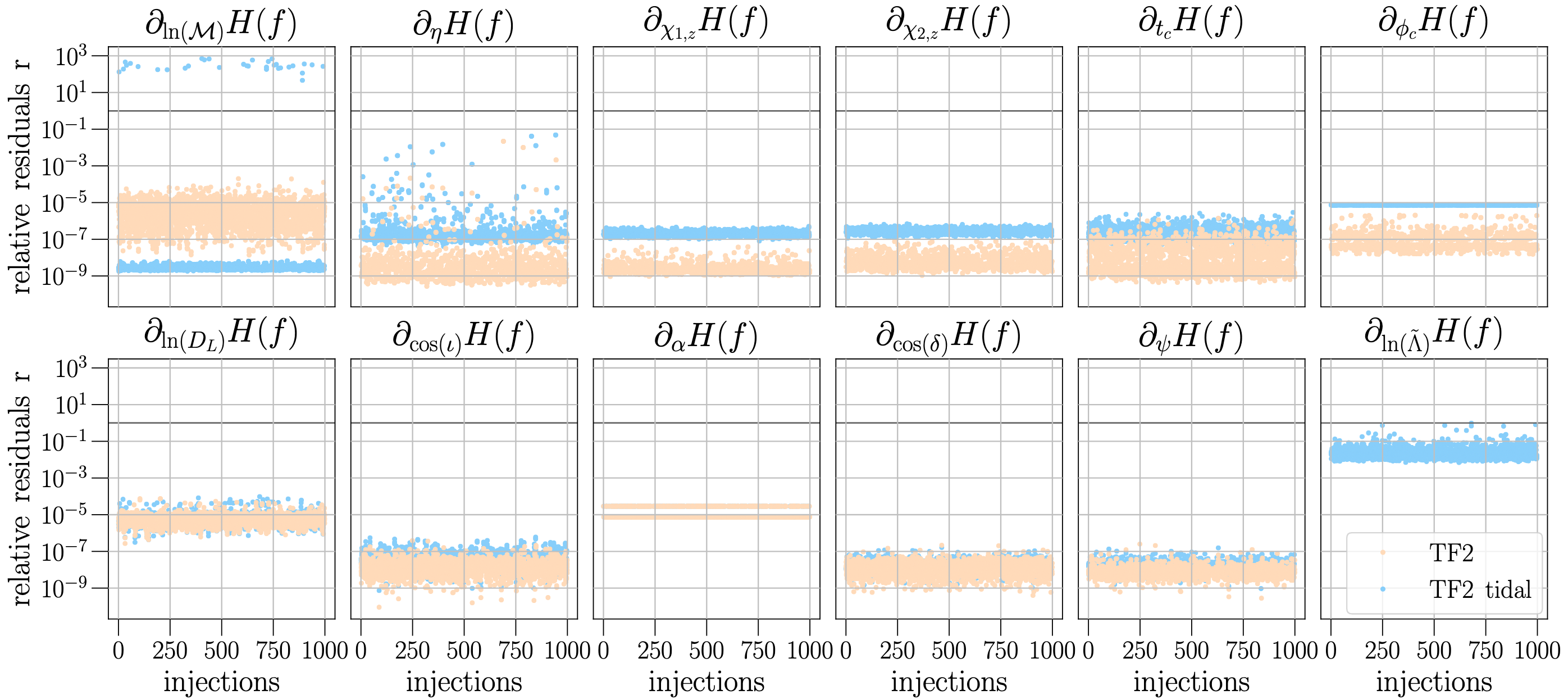}
    \caption{The plots show the relative residuals $r$, see Equation \eqref{eq:relative_res},
    between the numeric and symbolic partial derivatives of the detector responses for 11 (12) input
    parameters and WF model \texttt{tf2} (\texttt{tf2\_tidal}). The $\sim20$ very poor
    residuals in the chirpmass derivatives correspond to ill-conditioned Fisher matrices and are
    therefore disregarded for further analysis.}
    \label{fig:relative_res}
\end{figure*}

In the following, we present benchmarking results obtained with \gwb. We
compare numeric and symbolic derivatives and show their effects on the error
estimates. We further present error estimates from various WF models for
particularly interesting real events, namely GW170817 \cite{GW170817}, GW190521
\cite{GW190521}, and GW190814 \cite{GW190814}.

\subsection{Comparison of numeric and symbolic differentiation methods} \label{sec:num_vs_sym}

\gwbS contains both numeric and symbolic differentiation methods. The former can
be used with all WF models, while the latter only works with some of the
implemented WFs, as specified in Appendix \ref{app:wf_models}. In order to
compare the two approaches for the purposes of GW benchmarking, we will compute
the partial derivatives of the detector responses $H$ and the error estimates
for two WF models, TaylorF2 (\texttt{tf2}) and TaylorF2 with tidal effects
(\texttt{tf2\_tidal}) \cite{Lackey:2014fwa}.

Both WFs share the same set of input parameters, $(\mathcal{M}, \eta,
\chi_{1,z}, \chi_{2,z}, D_L, t_c, \phi_c, \iota, \alpha, \delta, \psi)$, with
the tidal WF further depending two tidal deformability parameter combinations
$(\tilde{\Lambda}, \delta\tilde{\Lambda})$ \cite{Wade:2014vqa}
\begin{align}
\tilde{\Lambda} &= \frac{8}{13}  \left [ (1 + 7  \eta - 31 \eta^2) \,(\Lambda_1 + \Lambda_2)\, +\right . \\
    & \phantom{= \frac{8}{13}} \left. \sqrt{1-4\eta} \, (1 + 9  \eta - 11  \eta^2)\, (\Lambda_1 - \Lambda_2)\right], \nonumber \\
\delta\tilde{\Lambda} &= \frac{1}{2\cdot1319} \cdot \\
    & \left [ \sqrt{1-4\eta} \,  (1319 - 13272  \eta + 8944 \eta^2) \,(\Lambda_1 + \Lambda_2)\, +\right . \nonumber\\
    & \left. (1319 -15910  \eta + 32850  \eta^2 + 3380 \eta^3)\, (\Lambda_1 - \Lambda_2)\right], \nonumber
\end{align}

\noindent where the component tidal deformabilities $\Lambda_1$ and $\Lambda_2$ are
defined as $\Lambda_i = \frac{2}{3}k_2C_i^5$ with Love number $k_2$,
compactness $C_i=\frac{m_i}{R_i}$, mass $m_i$, and radius $R_i$.

The partial derivatives are taken with respect to all available parameters
except $\delta\tilde{\Lambda}$ for the tidal WF. The detector responses are
evaluated for three detectors, each set to a 40km compact-binary optimized CE1
PSD, see Appendix \ref{app:tecs}.

We performed this comparison for a total of 1000 random sets of input
parameters for each WF. In both cases we chose the injections to be uniformly
distributed in co-moving volume up to a redshift of $z=0.5$ as well as over sky
positions and orientation angles. The masses and spins are also sampled
uniformly for both WFs albeit the ranges are chosen to resemble binary black
hole or binary neutron star systems, respectively: $5\,M_\odot \leq m_1,m_2
\leq 100\,M_\odot$ and $-0.75 \leq \chi_{1,z},\chi_{2,z} \leq 0.75$ for
\texttt{tf2} and $1\,M_\odot \leq m_1,m_2 \leq 2\,M_\odot$ and $-0.05 \leq
\chi_{1,z},\chi_{2,z} \leq 0.05$ for \texttt{tf2\_tidal}.

\begin{figure*}
    \includegraphics[width=\textwidth]{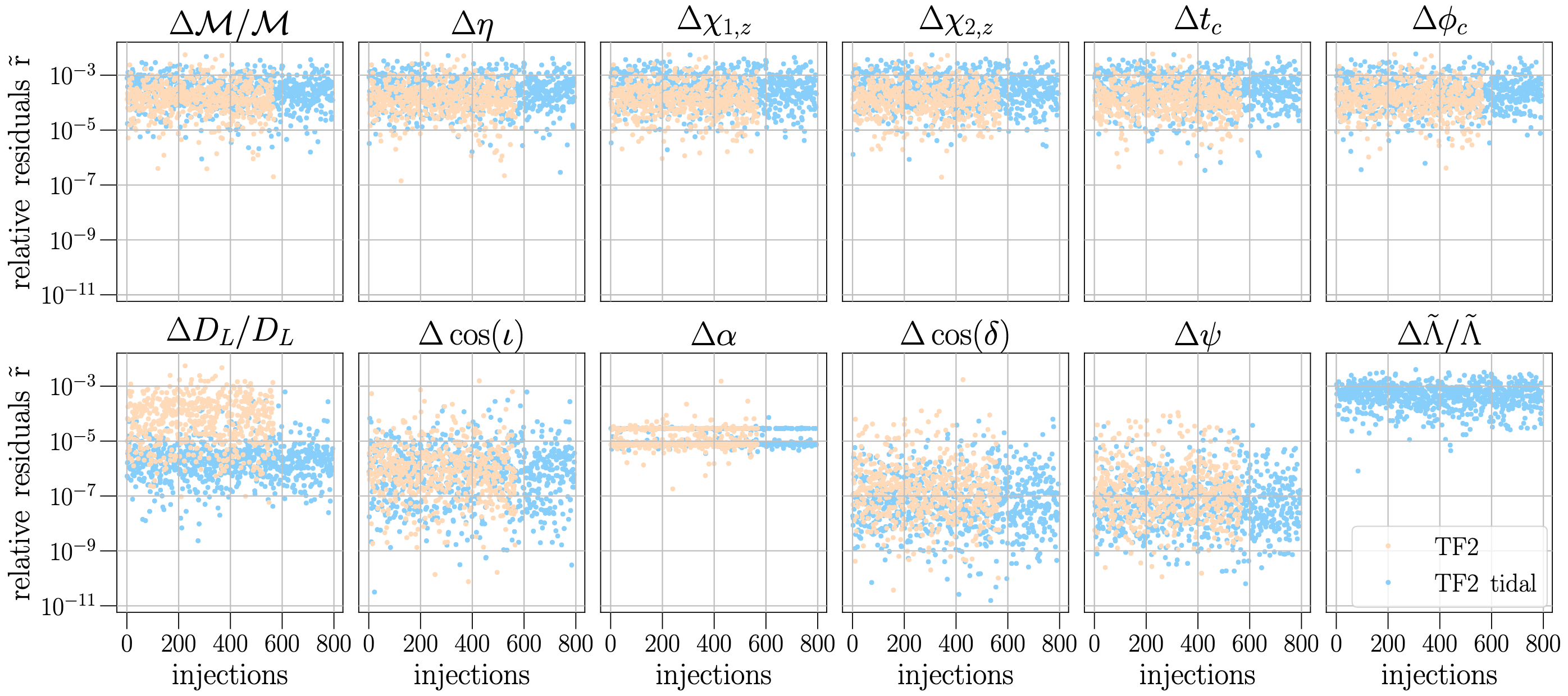}
    \caption{The plots show the relative residuals $\tilde{r}$, see Equation \eqref{eq:error_residuals},
    between the $1\sigma$-error bounds, estimated via the FIF from numeric or symbolic derivatives,
    for 11 (12) input parameters and WF model \texttt{tf2} (\texttt{tf2\_tidal}).}
    \label{fig:error_residuals}
\end{figure*}

Figure \ref{fig:relative_res} shows the relative residuals of all partial
derivatives of the detector responses for both waveforms and all three
detectors. The residuals are defined via
\begin{equation}
r_x = \frac{\sqrt{\sum_i (\partial_x^\text{num} H(f_i) - \partial_x^\text{sym}
H(f_i))^2}}{\sqrt{\sum_i (\partial_x^\text{sym} H(f_i))^2}} \label{eq:relative_res}
\end{equation}

\noindent where $i$ runs over the input frequency array and
$x\in\{\ln(\mathcal{M}), \eta, \chi_{1,z}, \chi_{2,z}, \ln(D_L), t_c, \phi_c,
\cos(\iota), \alpha, \cos(\delta), \psi, \ln(\tilde{\Lambda})\}$.

The inclusion of tidal effects results in two trends for the relative residuals
of the partial derivatives in Figure \ref{fig:relative_res}: The errors of the
derivatives with respect to the so-called intrinsic parameters, $(\mathcal{M},
\eta, \chi_{1,z}, \chi_{2,z}, t_c, \phi_c, \tilde{\Lambda},
\delta\tilde{\Lambda})$, worsen but are also more tightly constrained, while
the derivatives with respect to the extrinsic parameters $(D_L, \iota, \alpha,
\delta, \psi)$ appear to not be affected. The only exceptions are the chirpmass
and symmetric mass ratio derivatives, with the former showing improved
residuals with tidal effects and the latter showing more scatter.
The reason for the discrepancy in the behavior of the intrinsic and extrinsic
derivatives is found in the correlation of the intrinsic parameters to each
other, thus worsening the numeric derivatives when adding another intrinsic
parameter, whereas the extrinsic parameters stay unaffected. 

Overall, the partial derivatives appear to be well-behaved with relative
residuals well below $10^{-4}$. The tidal parameter derivatives perform the
worst, but stay below an error of 10\% for most of the samples. Finally, the
$\sim20$ very poor residuals in chirpmass derivatives actually correspond to
ill-conditioned Fisher matrices and are therefore disregarded for further
analysis.

Figure \ref{fig:error_residuals} shows the relative residuals
\begin{equation}
\tilde{r}_x = \left|\frac{\sigma_x^\text{num} - \sigma_x^\text{sym}}{\sigma_x^\text{sym}}\right|
\label{eq:error_residuals}
\end{equation}

\noindent of the $1\sigma$-errors bounds estimated from the FIF using the
numeric and symbolic derivatives of the detector responses.

While the relative residuals on the parameter error estimates from Figure
\ref{fig:error_residuals} do not show the actual discrepancy between the two
derivative methods, they do show their effect on the final product of the
Fisher analysis. It is important to note that the number of injections shown
has decreased due to some parameter combinations yielding ill-conditioned
Fisher matrices for which the further analysis is aborted. In fact, it appears
that the inclusion of tidal effects has a stabilizing effect for the FIF,
resulting in many more well-conditioned Fisher matrices. We remark that the
mentioned differences in the behavior of the residuals between \texttt{tf2} and
\texttt{tf2\_tidal} could be simply a symptom of the different input parameter
sets which we leave for a future study.

It appears that the correlations between intrinsic or extrinsic parameters
smudges the differences in the numeric and symbolic derivatives within each of
these parameter groups and the inclusion of tidal effects has little
influence on the residuals. The only exception is presented in the case of the
luminosity distance error bounds where the tidal WF resulted in improved errors.
Overall, the relative residuals between the parameter error bounds, estimated
with numeric and symbolic derivative methods, stay under 1\% and mostly under
0.1\%. Thus, we can use either approach confidently to perform the FIF.

\begin{figure*}
    \includegraphics[width=\textwidth]{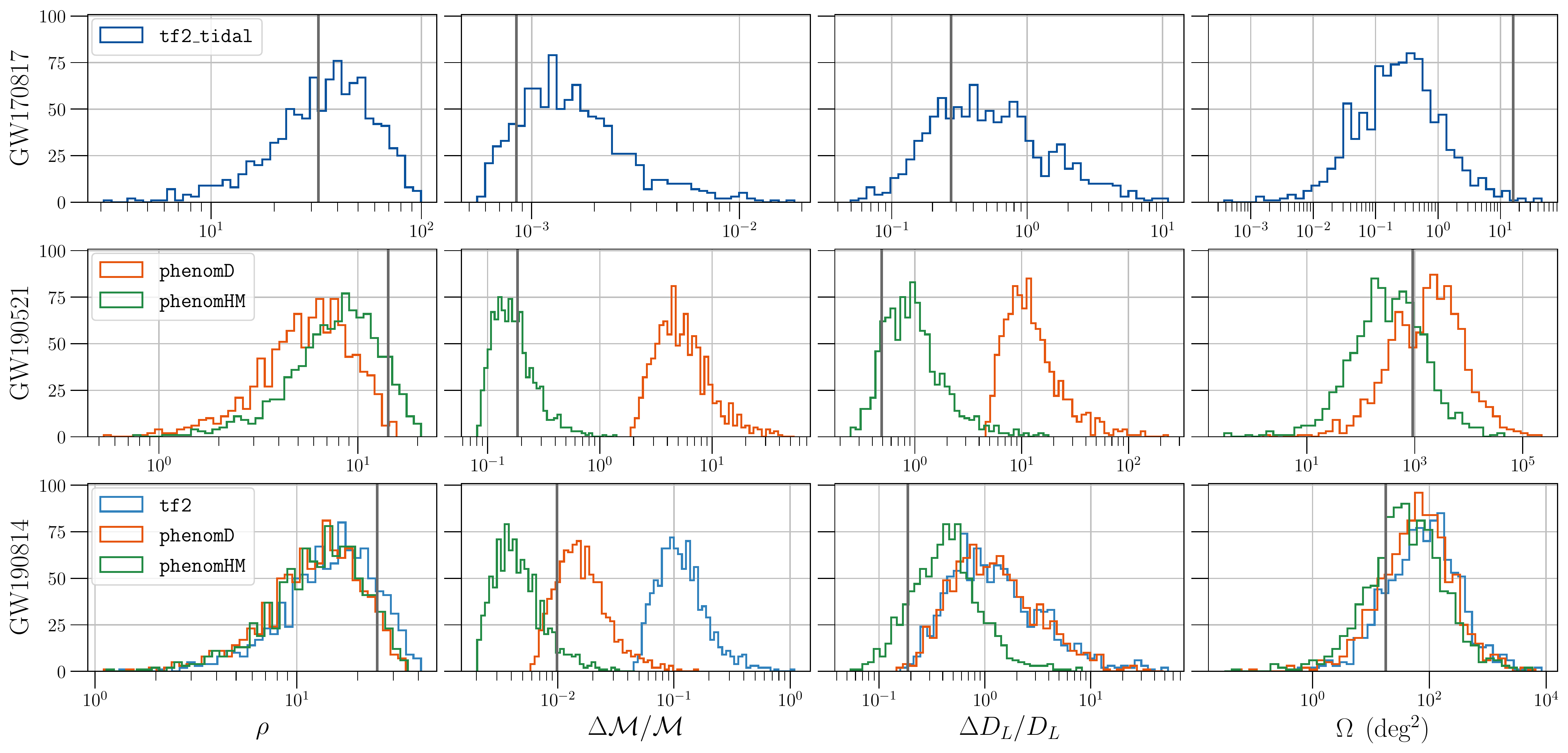}
    \caption{The figure shows the histograms of four quantities (SNR $\rho$, $1\sigma$-fractional
    error estimates of the chirpmass $\Delta \mathcal{M} / \mathcal{M}$ and luminosity distance
    $\Delta D_L / D_L$, and the 90\%-credible sky area $\Omega$) computed with \gwbS for three
    real events (GW170817, GW190521, and GW190814). The histograms are generated from 1000
    realizations of the sky positions and binary orientations while masses, spins, and
    luminosity distances are set to the measured values. The networks are set the same detector
    locations, LIGO Hanford, LIGO Livingston, and Virgo, and PSDs from the actual events. The
    vertical lines are the measured values from the respective events
    \cite{GWTC1,GW170817,GW190521}.}
    \label{fig:special_events}
\end{figure*}

\subsection{Application to real events}

GW170817 and GW190814 are among the loudest events observed by the LIGO and
Virgo detectors so far with SNRs $\rho_\mathrm{GW170817}=32$ and
$\rho_\mathrm{GW190814}=25$, respectively. These are exceptional candidates to
check the FIF and GW benchmarking against. GW190521 is another interesting event
representing signals of average loudness with $\rho_\mathrm{GW190521}=14$ and
very short duration as well as heavy binaries with merger- and
ringdown-dominated signals. Neither of these events have well determined sky
positions and binary orientations. We applied \gwbS to 1000 realizations of the
various angles while we fixed the masses, spins, and luminosity distances of the
systems to the reported median values \cite{GWTC1,GW170817,GW190521}. All three
events were detected by three detectors located at the LIGO Hanford, LIGO
Livingston, and Virgo sites. Since the actual sensitivities of these detectors
during the detections are determined by the facilities, we used the respective
PSDs for this analysis \cite{GWTC1_PSDs} and ran the FIF for 4 different WF
models: \texttt{tf2\_tidal} for GW170817, \texttt{IMRPhenomD}
\cite{Husa:2015iqa, Khan:2015jqa} and
\texttt{IMRPhenomHM} \cite{London:2017bcn} for GW190521, and \texttt{tf2}, \texttt{IMRPhenomD}, and
\texttt{IMRPhenomHM} for G190814.

We compare the reported values for the SNR $\rho$, the fractional error
estimates of the chirpmass $\Delta \mathcal{M} / \mathcal{M}$ and luminosity
distance $\Delta D_L / D_L$, as well as the 90\%-credible sky area $\Omega$ to
our benchmarking results. While this comparison contains only four quantities,
we still perform the FIF with respect to 11 (GW190521 and GW190814) or 12
(GW170817) parameters, in order to not underestimate the error bounds. These 11
and 12 parameters are the same sets as in Section \ref{sec:num_vs_sym} above for
\texttt{tf2} and \texttt{tf2\_tidal}, respectively. Besides, in order to allow
for an easier comparison with the measured values we converted the fractional
errors from the standard $1\sigma$-output of the FIF to the respective 90\%
credible bounds.

The benchmarking posteriors for the 1000 realizations are presented in Figure
\ref{fig:special_events} indicating that they capture the measured errors well
for at least one of the WF models per event. The figure shows that
\texttt{tf2\_tidal} performs really well for an inspiral dominated, binary
neutron star signal like GW170817. This does not hold for signals from the
heavier binaries.  For such systems, the error estimates from \texttt{tf2} do
not capture the measured values due to missed post-inspiral contributions
(GW190814), while the FIF completely failed for a strongly merger- and
ringdown-dominated signal (GW190521). In fact, for both of these systems it was
necessary to use \texttt{IMRPhenomHM}, a WF with higher modes, and not just
\texttt{IMRPhenomD} in order to fully capture the measured errors.

These findings are in line with the WFs used for the actual parameter
estimation analyses of these events and they show that \gwbS provides a
consistent tool for the purposes of GW benchmarking.

\section{Example applications} \label{sec:applications}

\gwbS has already been applied in two works:
\cite{Borhanian:2020vyr,Gupta:2020lxa}. In \cite{Borhanian:2020vyr} we examined
the possibility to use dark sirens---GW sources without an electromagnetic
counterpart---to resolve the tension in the measurement of the Hubble constant.
In \cite{Gupta:2020lxa} we studied how multiband observations of GW signals
with the space-based LISA and ground-based 3G observatories will open up new
possibilities to conduct multiparameter tests of general relativity in the line
of \cite{Arun:2006hn}.

\subsection{``Dark sirens to resolve the Hubble-Lema\^{i}tre tension''}

We examined the prospect to use dark sirens to perform precision measurements
of the Hubble constant and thus provide a GW-assisted method to resolve the
Hubble-Lema\^{i}tre tension. The proposed method makes use of the potential of
future observatories to measure sky positions of close-by dark sirens well
enough to pin-point the host galaxy of the GW event. This provides two
independent measurements, the distance from the GW signal and the galaxy's
redshift from electromagnetic follow-up observations, which compound to
estimate the Hubble constant.

We simulated three populations of dark sirens: one resembled binary black holes
from the GWTC-1 catalog \cite{GWTC1}, the second focused on the heavy
sub-population of such BBHs with component masses $m_1,m_2>25\,M_\odot$, while
the third was GW190814-like \cite{GW190814}. The GWTC-1 catalog summarizes the
detections from the first two observing runs of the LIGO and VIRGO detectors
and GW190814 was the most asymmetric binary observed to date.

For each population we performed the Fisher analysis in four networks
containing three possible detector technologies of the future ('Advanced+',
'Voyager', and 3G). In order to leverage the sensitivities of such future
networks we used the \textsc{lalsimulation} WF \texttt{IMRPhenomHM}
\cite{London:2017bcn} which also models subdominant, higher-order spherical
harmonic modes beyond the quadrupole. These higher modes improve the distance
measurement by breaking the degeneracy between the luminosity distance and the
inclination angle of the binary orbit to the line of sight, thus ultimately
yielding better measurements of the Hubble constant.

Using \gwbS we performed the FIF with numeric derivatives for $10^4$
binaries per population. We estimated the number of detections to expect
from the studied networks that yield precise measurements of the dark siren's
sky position and propagated the error bounds on the luminosity
distance measurement to estimate the accuracy of a Hubble constant measurement.
We showed that, if redshift errors are well controlled, dark siren observations will 
yield $\lesssim2\%$-accurate Hubble constant measurements 
already in the Advanced+ era and thus resolve the Hubble-Lema\^{i}tre tension.

\subsection{``Multiparameter tests of general relativity using multiband gravitational-wave
observations''}

Parameterized post-Newtonian tests \cite{Arun:2006hn} present a crucial
resource to perform theory-agonistic tests of general relativity which can
probe generically potential deviations of GW signals from Einstein's theory.
This class of tests is especially important when verifiable GW predictions from
alternate theories of gravity are sparse. The tests are based on modifications
to the expansion coefficients in the post-Newtonian approximation to the WFs
emitted by compact binary coalescences, such that these deviations vanish if
general relativity is correct. Unfortunately, the current tests are handicapped
by the strong correlations between the introduced deviation parameters.
Therefore, parameterized post-Newtonian tests are limited to single-coefficient
modifications for now. We showed that the combination of the information
present in the low-frequency LISA and high-frequency 3G bands breaks this
degeneracy. Thus such multiband observations allow for true multiparameter
tests of general relativity.

In our study we examined stellar mass binary black hole coalescences. Their
inspiral signals will last several years in the LISA band ($\sim
0.1-100\,\mathrm{mHz})$ before they reach the 3G band ($\sim$ Hz-kHz) where
they finally merge. Unfortunately, these signals, while extremely loud in 3G
networks, will be faint to non-visible in LISA, making the `joint'
observation a difficult task. The issue lies in the detection methodology for
such quiet signals. The current state-of-the-art are blind, matched filtering
searches \cite{gstlal_live,pycbc_live} without a-priori knowledge of the GW
signal. Such blind searches require GW template banks which cover a large
volume in the parameter space. Thus, the sizes of the banks become
computationally infeasible for the signals of interest in the LISA band
\cite{Moore:2019pke}.

Fortunately, we can leverage the high-fidelity detections with 3G detectors to
dig out these quiet signals from LISA's archival data. In contrast to a blind
search the 3G detection confirms the existence of the signal we are searching
for and further provides information about it, allowing for the generation
of signal-specific template banks that decrease the computational cost
immensely. This is enabled in two-ways: A 3G network will estimate all WF
parameters except chirpmass and symmetric mass ratio to better precision than
LISA could, thus decreasing the parameter dimension of the template bank to two
and fixing the other parameters to the 3G measurement. The search volume can be
further narrowed down to within the error bounds estimated from the 3G
detection for both chirpmass and symmetric mass ratio.

Using \gwb, we tackled the question of feasibility to search the archival LISA
data following a 3G detection, strengthening our proposal to perform multiband
multi-parameter tests of general relativity. For this purpose we generated a
set of 500,000 binary black holes up to a redshift of $z=10$. Roughly 200 of
these binaries should be visible in LISA with an SNR $\geq 4$ which we set as
the threshold for visibility in the archival search. \gwbS allowed us to
estimate the SNR and $1\sigma$ error bounds on the WF parameters for a network
consisting of one Einstein Telescope and two $40\,\text{km}$, CE2 detectors in
compact binary optimization.

Given the LISA SNR $\rho_\text{LISA}$ and 3G errors on chirpmass $\sigma_{M_c}$
and symmetric mass ratio $\sigma_\eta$, we used \gwbS to calculate the
detector-dependent volume $V$ via the metric $g$ \cite{Owen:1995tm,Owen:1998dk}
\begin{equation}
g = \rho_\text{LISA}^{-2} \det\left(\langle\partial_x H,\partial_y H\rangle_{x,y\in\{M_c,\eta\}}\right)
\end{equation}

\noindent as
\begin{equation}
V = \int\limits_{M_c-2\sigma_{M_c}}^{M_c+2\sigma_{M_c}}\md M_c \int\limits_{\eta-2\sigma_{\eta}}^{\eta+2\sigma_{\eta}} \md \eta \, \sqrt{g}.
\end{equation}

\noindent The inner product in the definition of the metric is taken with
respect to the LISA PSD.  Finally, the number of templates $N$ is given by the
ratio of the total volume to the volume of the error ball around each template
for a given minimal match $mm$
\begin{equation}
N = \frac{V}{1-mm}.
\end{equation}

Using $mm=0.95$, we found that such targeted searches of LISA's archival data
following the detection of a binary black hole merger in a network of 3G
detectors should allow for template bank sizes of the order $10^3$ -- $10^4$.
This is substantially less than the value of $10^{12}$ predicted before
\cite{Moore:2019pke}.  Ultimately, \gwbS enabled us to check the feasibility of
multibanding to perform multi-parameter tests of general relativity by
providing the tools to calculate estimates of both the 3G errors bounds and the
LISA template numbers.


\section{Summary}\label{sec:summary}

In this work we presented \gwb, a \textsc{Python} package for fast and
straightforward applications of the Fisher information formalism for the
purposes of GW benchmarking. The package is written with the usage for detector
networks in mind. Thus, it is structured around the \networkS class which acts
as a hub for user interaction and facilitates all the possible computations.
The package gives easy access to a variety of GW benchmarking quantities such
as the frequency-domain WF polarizations, antenna pattern functions and
location phase factors as well as noise PSDs for various detector locations and
technologies, frequency-domain detector responses, SNRs, Fisher and covariance
matrices, and the product of the FIF, the $1\sigma$ error bounds for input
parameters of choice.

There are two types of WFs available, both of which support numeric differentiation
via the \textsc{numdifftools} package: \textsc{LAL} WFs as well as those
implemented \gwb. The latter further support symbolic derivatives via
\textsc{sympy} which are faster and more accurate. The inclusion of \textsc{LAL}
WF support is very important as it gives access to a host of WF models that are
well tested and commonly used in the literature. \gwbS contains nine existing,
planned, or fiducial detector locations and 23 different detector technologies
ranging from aLIGO to various CE sensitivity curves. The addition of
user-defined locations and technologies is also possible. Another specialty of
this package is the inclusion of the effects of Earth's rotation in the antenna
patterns and detector responses, and thus the FIF. This is particularly
important for long-lived GW signals such as from binary neutron stars.

\gwbS is and has been used for several research projects of which we presented
two that are in public already. In both cases, \gwbS allowed us to generate and
calculate parameter errors and even template numbers for a matched filtered
search for large numbers of sources and detector networks.  \gwbS represents a
unique code that implements the FIF and GW benchmarking in an easy-to-use
manner for a wide range of problems, WFs, and network options and allows for
straightforward application in large-scale simulations for improved statistics
of the scientific claims.

\acknowledgements
I thank Bangalore Sathyaprakash, Anuradha Gupta, Patrick Godwin, and everyone in
the Cosmic Explorer Project for helpful comments and discussions during the
development of \gwb. I further thank Kevin Kuns, Varun Srivastava, Evan Hall,
Matthew Evans, and Stefan Ballmer for providing the sensitivity curves used in
\gwb. I also thank all front line workers combating the CoVID-19 pandemic and
those maintaining ordinary life without whose support this work would not have
been possible. I acknowledge the support by NSF grant PHY-1836779.  Computing
resources for the development and testing of this code were provided by the
Pennsylvania State University.

\appendix

\section{\networkS and \detectorS classes}\label{app:net_vs_det}

GW detector networks consist, as per name, of a number of detectors that work in tandem to observe
GWs to improve signal strength, sky coverage, and parameter estimation. Leaning on
this understanding, the \networkS class is designed as a GW benchmarking hub for a chosen set
of detectors. The class structure allows for straightforward handling of the relevant quantities and
network methods and thus provides the user with tools to load, calculate, and manipulate quantities
like waveforms, SNRs, or errors estimates.

The \detectorS class is written with similar syntax and methodology as the \networkS class. This is
very natural because network quantities like SNR or error estimates are handled in the same
manner as the respective single detector ones. This choice allows for clearer codes and faster
development, but also better understanding of both classes, which helps when using \networkS
objects. Thus, Table~\ref{tab:net_vs_det} compares the \emph{instance variables} of both classes
side-by-side. The top section of the table lists the variables which are set by the user and tell
the \network

\begin{itemize}
\item the label of the \networkS and what location and technology to use for a \detectorS instance,
\item which \detectorS instances to initialize,
\item what frequency array and injection parameters to use for function evaluation, and with
respect to which parameters to take derivatives for error estimation via FIF,
\item what \waveform to initialize,
\item which parameters $x$ to convert to $\cos(x)$ or $\ln(x)$ to estimate errors on the latter,
and whether to take Earth's rotation into account.
\end{itemize}

The bottom section shows the variables which can be loaded and computed with instance methods
based on the user-set variables:

\begin{itemize}
\item Detector-specific quantities like PSDs, antenna patterns, and location phase factors are
loaded and stored in each \detectorS instance.
\item The WF polarizations and the respective derivative dictionaries are detector-agnostic and
saved in the \network.
\item Each \detectorS handles the respective detector response and derivative dictionaries.
\item The detector and network SNRs (and their squares) are stored in the respective instances,
where each \detectorS also saves the integrand of the square SNR.
\item Each detector and the network contain their respective copy of the FIM, its condition number,
a well-conditioned flag, the covariance matrix, the inversion error, and an error dictionary
containing the $1\sigma$ errors extracted from the covariance matrix.
\end{itemize}


\section{Implemented waveform models and the \waveform class} \label{app:wf_models}

\begin{table*}
\centering
    \caption{\networkS vs \detectorS variables.}
    \begin{tabular}{p{0.2\linewidth}|p{0.39\linewidth}|p{0.39\linewidth}}
    \hline \hline
    & \network & \detector \\
    \hline \hline
    \multicolumn{3}{c}{\multirow{2}{*}{\emph{set by user:}}} \\
    \multicolumn{3}{c}{} \\
    \hline
    labels & \texttt{.label}~~~(label to initialize network) & \texttt{.det\_key, .tec, .loc}~~~(full detector key, technology and location labels) \\
    \hline
    detectors & \texttt{.detectors, .det\_keys}~~~(list of \detectorS objects and detector keys) & --- \\
    \hline
    global quantities & \texttt{.f, .inj\_params, .deriv\_symbs\_string, .deriv\_variables}~~~(frequency array, injection parameters, derivative parameters string and list)
    & \texttt{.f}~~~(truncated frequency array) \\
    \hline
    waveform & \texttt{.wf}~~~(\waveform object) & --- \\
    \hline
    analysis settings & \texttt{.conv\_cos, .conv\_log, .use\_rot}~~~(lists
    of parameters $x$ to convert to $\cos(x), \ln(x)$, consider Earth's rotation) & --- \\
    \hline
    \multicolumn{3}{c}{\multirow{2}{*}{\emph{loaded or calculated by network methods:}}} \\
    \multicolumn{3}{c}{} \\
    \hline
    detector specific quantities & --- & \texttt{.psd, .Fp, .Fc, .Flp}~~~(detector PSD, antenna patterns, location phase factor) \\
    \hline
    waveform polarizations & \texttt{.hfp, .hfc, .del\_hfpc, .del\_hfpc\_expr}~~~(plus and cross
    polarizations, derivatives dictionary, \textsc{sympy} expressions dict)
    & --- \\
    \hline
    detector responses & --- & \texttt{.hf, .del\_hf, .del\_hf\_expr}~~~(detector response, derivatives dictionary, \textsc{sympy} expressions dict) \\
    \hline
    SNR & \texttt{.snr, .snr\_sq}~~~(SNR, SNR square) & \texttt{.snr, .snr\_sq, .d\_snr\_sq}~~~(SNR, SNR square, SNR square integrand)\\
    \hline
    error quantities & \texttt{.fisher, .cond\_num, .wc\_fisher, .cov, .inv\_err, .errs}~~~(Fisher information matrix, Fisher condition number, well conditioned boolean, covariance matrix, inversion error, the parameter errors)
    & \texttt{.fisher, .cond\_num, .wc\_fisher, .cov, .inv\_err, .errs}~~~(Fisher information matrix, Fisher condition number, well conditioned boolean, covariance matrix, inversion error, the parameter errors) \\
    \hline \hline
    \end{tabular}
    \label{tab:net_vs_det}
\end{table*}

\begin{table*}
    \centering
    \caption{The available WFs are presented here, together with their labels, their input
    parameters, and whether they allow for numeric or symbolic differentiation.
    \gwbS uses the same labeling convention for \texttt{'approximant'} in the \textsc{LAL} WFs as specified in
    \textsc{lalsimulation}. Check the \emph{Approximant}-list for
    frequency-domain approximants, implemented in \texttt{lalsimulation}, under \emph{Enumerations} at
    \url{https://lscsoft.docs.ligo.org/lalsuite/lalsimulation/group___l_a_l_sim_inspiral__h.html}.
    }
    \begin{tabular}{p{0.13\textwidth}|p{0.13\textwidth}|p{0.42\textwidth}|p{0.25\textwidth}}
    \hline
    \hline
    Waveforms & Waveform label & Input parameters & Other parameter dictionary \\
    \hline
    \hline
    \multicolumn{4}{c}{\multirow{2}{*}{\emph{\textsc{LAL} WFs --- only numeric differentiation}}} \\
    \multicolumn{4}{c}{} \\
    \hline
    \hline
    BBH & \texttt{'lal\_bbh'} & \texttt{'f Mc eta chi1x chi1y chi1z chi2x chi2y chi2z DL tc phic iota'} & \texttt{\{'approximant':name\_string\}}\\
    \hline
    BNS & \texttt{'lal\_bns'} & \texttt{'f Mc eta chi1x chi1y chi1z chi2x chi2y chi2z DL tc phic iota lam\_t delta\_lam\_t'} & \texttt{\{'approximant':name\_string\}}\\
    \hline
    \hline
    \multicolumn{4}{c}{\multirow{2}{*}{\emph{Coded WFs --- numeric and symbolic differentiation}}} \\
    \multicolumn{4}{c}{} \\
    \hline
    \hline
    TaylorF2  & \texttt{'tf2'} & \texttt{'f Mc eta chi1z chi2z DL tc phic iota'} & None \\
    \hline
    TaylorF2 + tidal  & \texttt{'tf2\_tidal'} & \texttt{'f Mc eta chi1z chi2z DL tc phic iota lam\_t delta\_lam\_t'} & None \\
    \hline
    \hline
    \end{tabular}
    \label{tab:wfs}
\end{table*}

\gwbS handles WF selection and evaluation within the \waveform class. The
available WFs are listed in Table \ref{tab:wfs} together with their labels,
their input parameters, and whether they allow for numeric or symbolic
differentiation. One of the strong suites of \gwbS is the implementation of WF
models from \textsc{LAL} in the \waveform class and the capability to use them
for the FIF with numeric derivatives. The class contains five instance variables

\begin{itemize}
        \item \texttt{.wf\_model\_name}: internal name of the WF model,
        \item \texttt{.wf\_other\_var\_dic}: extra input parameters,
        \item \texttt{.wf\_symbs\_string}: string containing the basic
        parameters (space-separated),
        \item \texttt{.hfpc\_np}: \textsc{numpy} function,
        \item \texttt{.hfpc\_sp}: \textsc{sympy} function,
\end{itemize}

\noindent and few methods of which two are of particular interest to the user:

\begin{itemize}
        \item \texttt{.get\_sp\_expr()}: load the \textsc{sympy} function to obtain a symbolic
        expression which can be manipulated,
        \item \texttt{.eval\_np\_func(f,inj\_params)}: evaluates the \textsc{numpy} function for the
        given frequency array \texttt{f} and injection parameters \texttt{inj\_params}. \\
        (The injection parameters \texttt{inj\_params} can be either provided as a dictionary with keys equal
        to the parameters defined in \texttt{.wf\_symbs\_string} or as a list with the same ordering
        given by \texttt{.wf\_symbs\_string}.)
\end{itemize}

\begin{table*}
\centering
    \caption{The available detector locations are presented together with the respective labels and detector
    angles (longitude $\alpha_d$, latitude $\beta_d$, and orientation of the y-arm with respect to due
    East $\gamma_d$). The five locations H, L, V, K, I are chosen in accordance
    with the implementation in \textsc{LAL}
    (\url{https://lscsoft.docs.ligo.org/lalsuite/lal/_l_a_l_detectors_8h_source.html}). The four locations E, C, N, and S are fiducial example sites and do not represent candidates under active consideration.}
    \begin{tabular}{c|l|c|c|c}
    \hline \hline
    \multirow{2}{*}{Label} & \multirow{2}{*}{Location} & \multicolumn{3}{c}{angles ($\text{rad}$)} \\
    \cline{3-5}
    & & $\alpha_d$ & $\beta_d$ & $\gamma_d$  \\
    \hline \hline
    H & LIGO Hanford (Washington, USA)                      & $-2.084060$ &  $0.810795$ & $-2.513290$  \\
    \hline
    L & LIGO Livingston (Louisiana, USA)                    & $-1.584310$ &  $0.533423$ & $-1.261590$  \\
    \hline
    V & Virgo (Cascina, Italy)                              & $ 0.183338$ &  $0.761512$ & $ 2.802430$  \\
    \hline
    K & KAGRA (Kamioka, Japan)                              & $ 2.394200$ &  $0.632682$ & $ 2.087480$  \\
    \hline
    I & LIGO India (Hingoli, India)                         & $ 1.334013$ &  $0.248418$ & $ 1.570800$  \\
    \hline
    E & Einstein Telescope (Cascina, Italy)                 & $ 0.183338$ &  $0.761512$ & $ 2.802430$  \\
    \hline
    C & Cosmic Explorer (main) (Idaho, USA)                 & $-1.969170$ &  $0.764918$ & $ 0.000000$  \\
    \hline
    N & Cosmic Explorer North (New Mexico, USA)             & $-1.858430$ &  $0.578751$ & $-1.047200$  \\
    \hline
    S & Cosmic Explorer South (New South Wales, Australia)  & $ 2.530730$ & $-0.593412$ & $ 0.785398$  \\
    \hline \hline
    \end{tabular}
    \label{tab:locs}
\end{table*}

Only the \texttt{wf\_model\_name} and \texttt{wf\_other\_var\_dic} are needed to
fully initialize a \waveform instance:

\begin{lstlisting}
from gwbench import wf_class

wf = wf_class.Waveform(
    wf_model_name = wf_model_name,
    wf_other_var_dic = wf_other_var_dic)
\end{lstlisting}

The \textsc{LAL} WF models do not provide any of the aforementioned symbolic
functionality, enabled by \textsc{sympy}, as this requires a direct implementation
of the WF model to ensure proper \textsc{sympy} function calls.


\section{Implemented detector locations - antenna patterns and location phase factor} \label{app:locs}

\begin{figure}[hb]
    \includegraphics[width=0.5\textwidth]{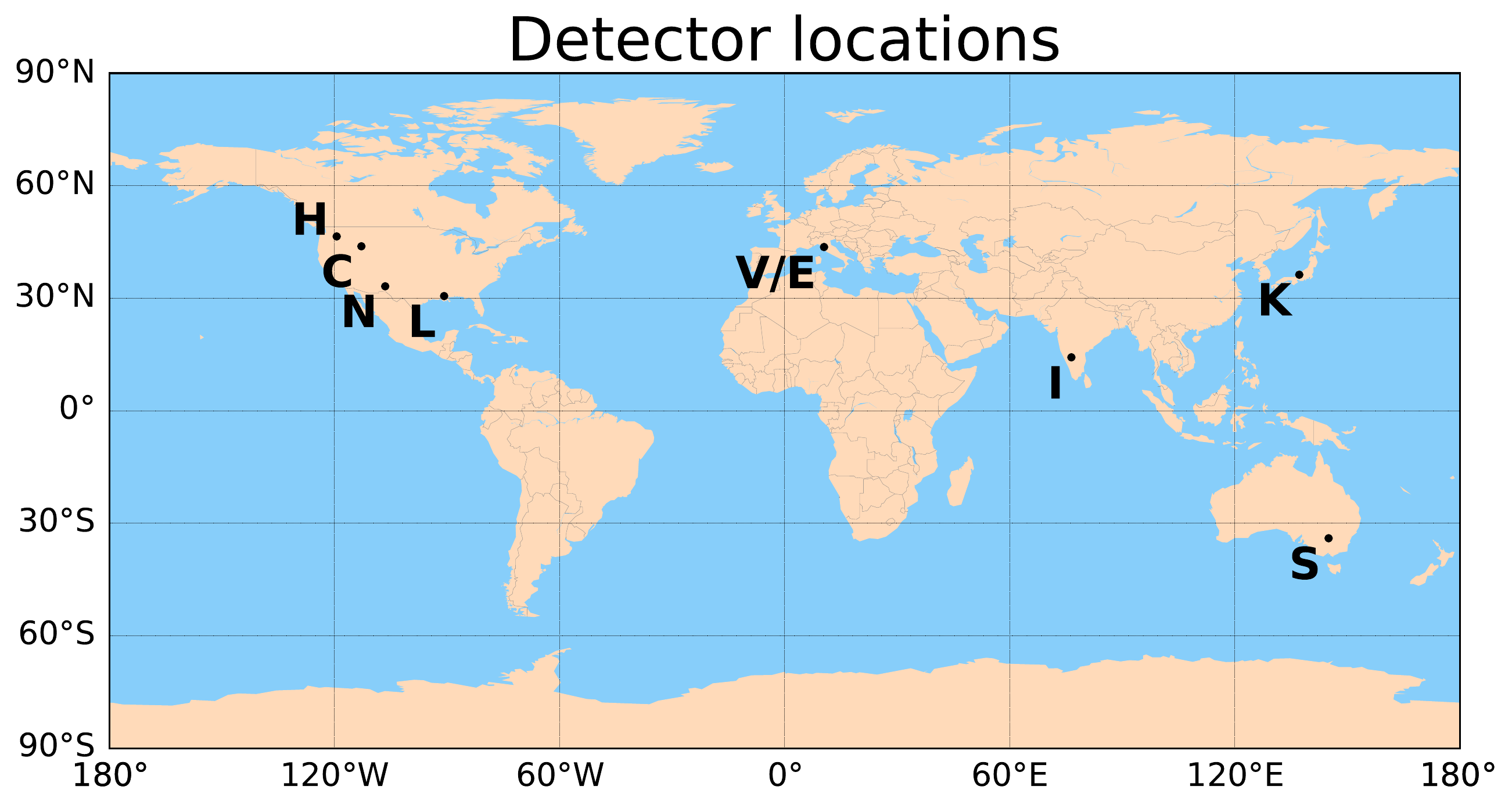}
    \caption{The available detector locations shown on the world map.}
    \label{fig:locs_map}
\end{figure}

Table \ref{tab:locs} summarizes the nine available detector locations in \gwbS
together with the respective labels used in the code, while Figure
\ref{fig:locs_map} presents their spread across the world map. There are five 2G
detector sites at LIGO Hanford (Washington, USA), LIGO Livingston (Louisiana,
USA), VIRGO (Cascina, Italy), KAGRA (Kamioka, Japan), and LIGO India (Hingoli,
India) as well as four fiducial sites for the 3G detectors. The Einstein
Telescope is set to the same coordinates as VIRGO, while the three Cosmic
Explorer locations Main, North, and South are set to sites in Idaho (USA), New
Mexico (USA), and New South Wales (Australia)\footnote{The four fiducial 3G
detector locations are example sites and do not represent candidates under
active consideration.}.

\begin{figure*}
    \includegraphics[width=\textwidth]{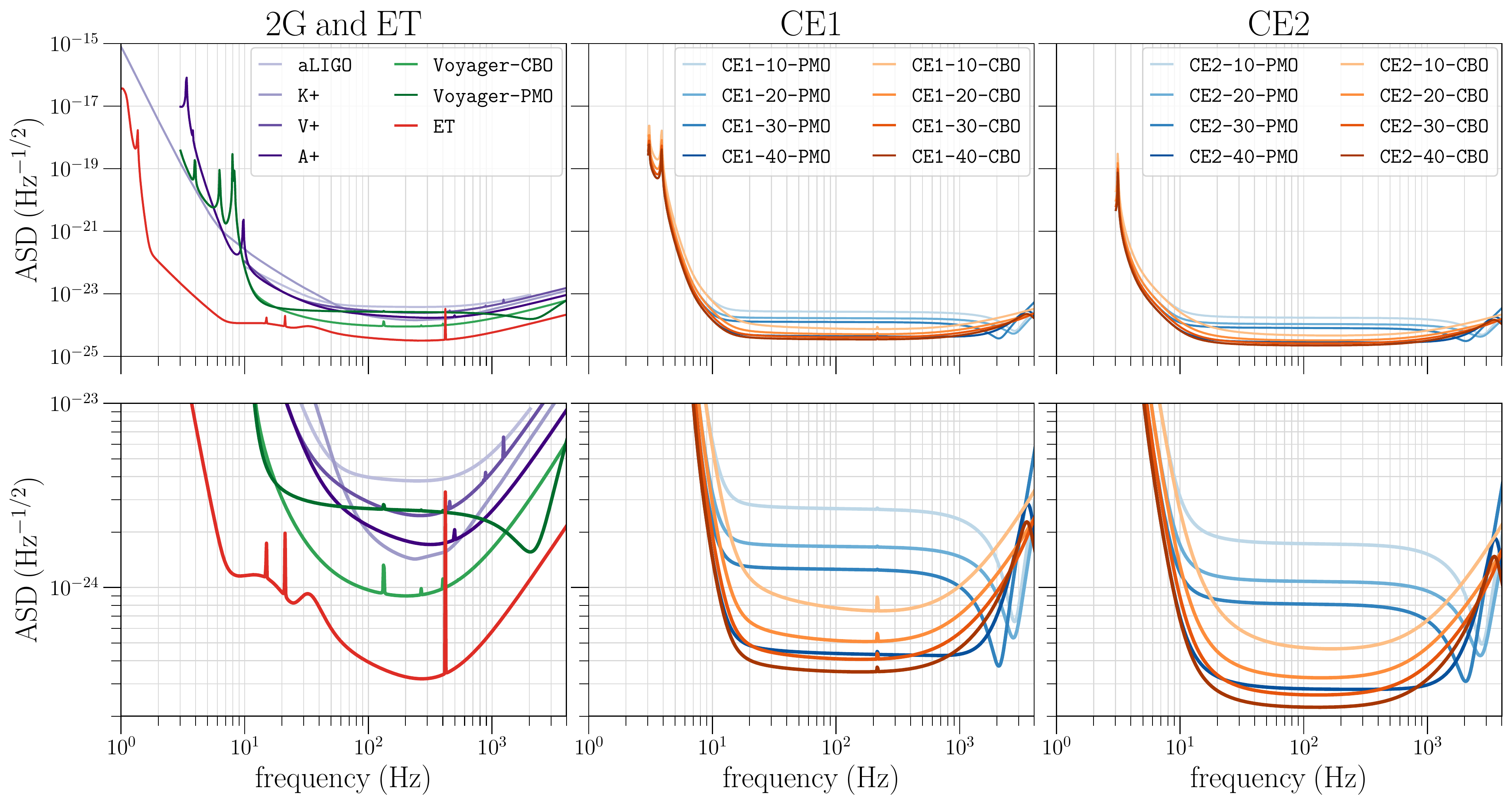}
    \vspace{10pt}
    \includegraphics[width=\textwidth]{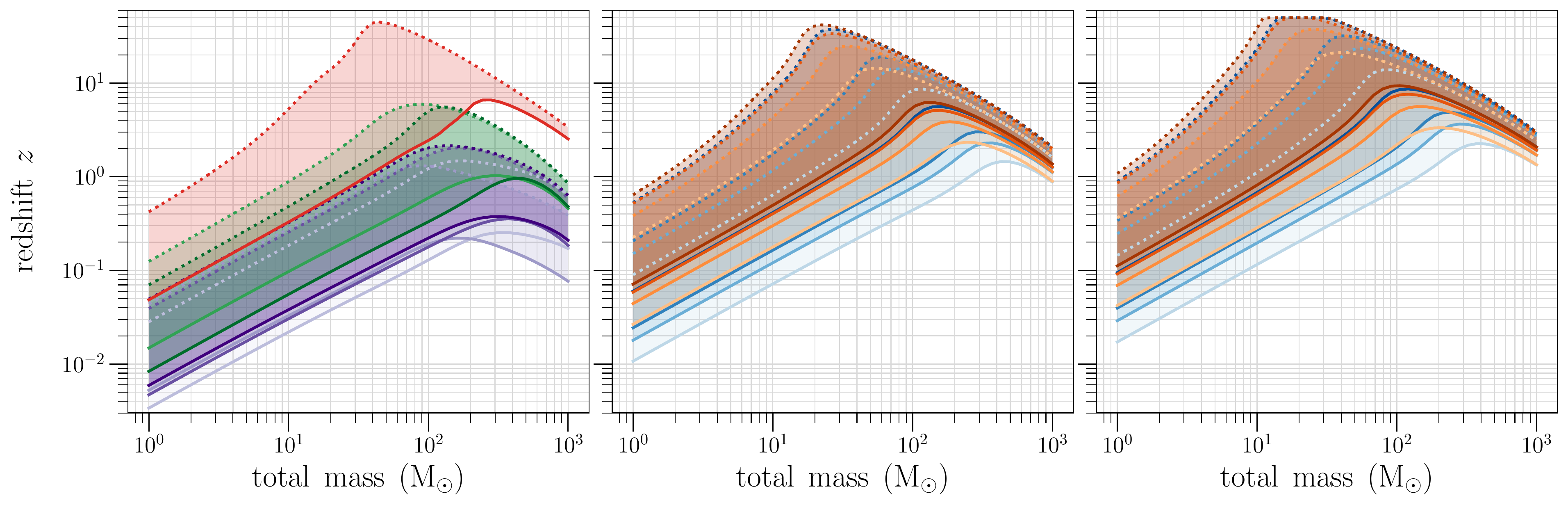}
    \caption{The \emph{top} shows the amplitude spectral densities (ASD) $\sqrt{S_n(h)}$ available in \gwb, see Figure \ref{fig:psds},
    as functions of the frequency between 1 and 4096 Hz. The \emph{bottom} presents the observable redshifts for equal-mass binaries
    at SNR 100 (solid) and 12 (dotted) as functions of the total binary mass between 1 and 1000
    $M_\odot$ (in the source frame). The systems are assumed to be optimally orientated for an L-shaped detector with the
    respective ASD. The SNRs were obtained for the \textsc{LAL} WF \texttt{IMRPhenomHM}. The shaded regions show the respective ranges for events with SNRs between 12 and 100.}
    \label{fig:psd_plots}
\end{figure*}

The computation of the antenna patterns $F_+,F_\times$  and location phase factors $F_{lp}$ for these
locations is implemented in the two modules \textsc{antenna\_pattern\_np.py} and
\textsc{antenna\_pattern\_sp.py} which differ in their intended usage: the former is used to
calculate the respective quantities directly whereas the latter is required for the generation of
the symbolic derivatives via \textsc{sympy}.

The core concepts for a single detector are demonstrated in
\cite{LRR_Sathyaprakash_2009}, whereas \cite{Schutz:2011tw} outlines the more
intricate methods required for the analysis of a network of detectors where
calculations are carried out in a universal frame centered at Earth's center. In
either case, $F_+$, $F_\times$, and $F_{lp}$ depend on a few signal-specific
quantities: right ascension $\alpha$, declination $\delta$, and polarization
angle $\psi$. In a network, these functions further depend on the arrival time
$t_E$ of the signal at Earth's center as well as three detector- and
location-specific angles: the longitude $\alpha_d$ and latitude $\beta_d$ of the
detector site as well as the angle $\gamma_d$ of the detector's y-arm with
respect to due East\footnote{Since the Einstein Telescope consists of three
V-shaped detectors ET1, ET2, and ET3, that form a equilateral triangle,
$\gamma_d$ is fiducially set to correspond to ET1. The fixed geometry allows the
calculation of the detector quantities for all three detectors.}.

All these considerations omit the effects of Earth's rotation on the antenna
patterns and location phases factors. These effects are negligible for
short-lived, transient signals lasting only a few minutes or less. They cannot
be neglected for 3G detectors which will see signals that can be up to a few
days long. Hence, \gwbS contains the functionality to take Earth's rotation,
simplified as a constant circular rotation with a period of one day, into
account which follows the calculations detailed in
\cite{Wen:2010cr,Zhao:2017cbb}.


\section{Implemented detector technologies - detector PSDs} \label{app:tecs}

\begin{figure}[ht]
    \includegraphics[width=0.5\textwidth]{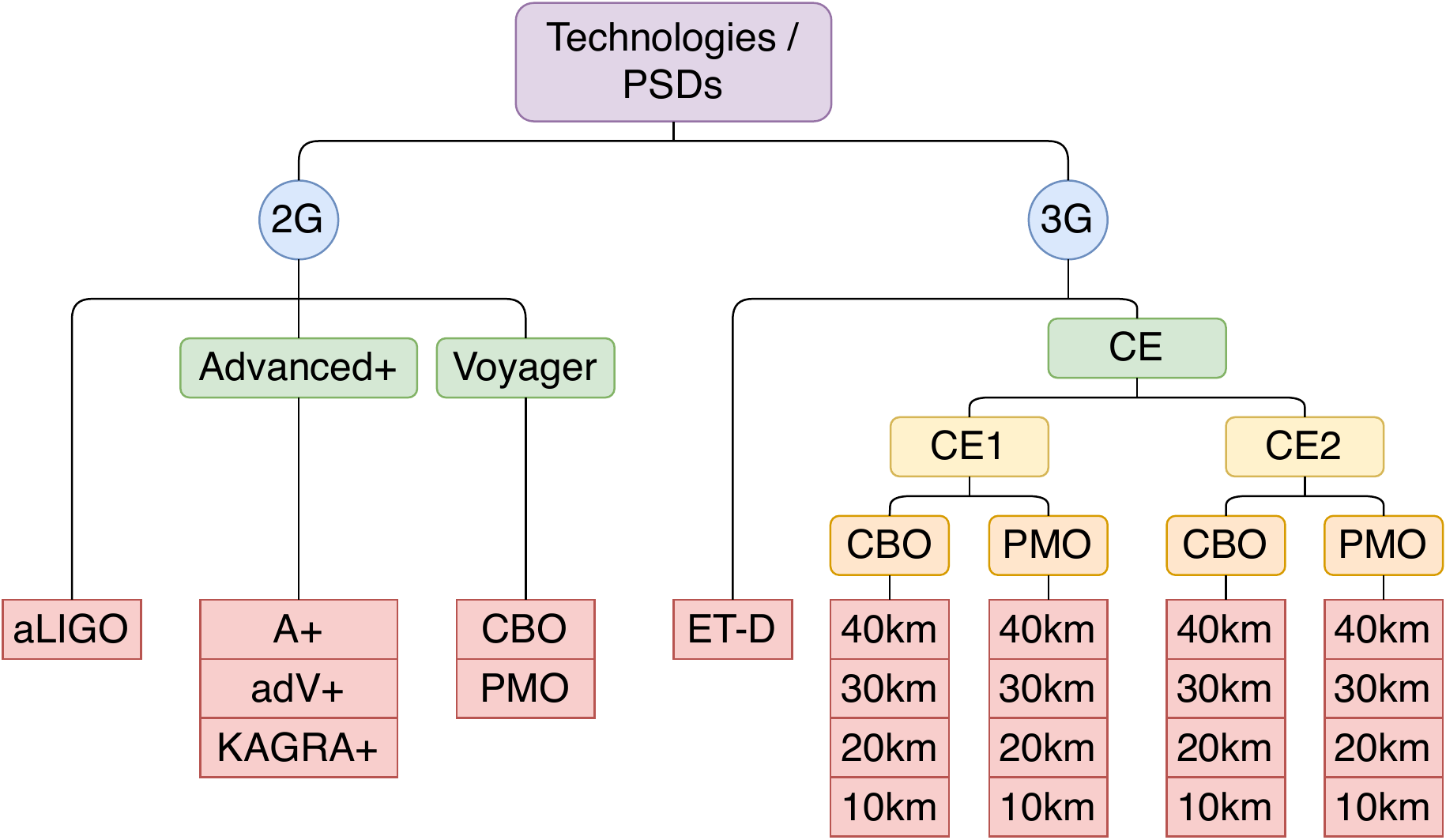}
    \caption{Detector technologies/PSDs available in \gwb. The included 2G PSDs
    encompass aLIGO as well as the planned Advanced+ and proposed Voyager
    upgrades. The 3G PSDs encompass the ET-D curve for the Einstein Telescope
    and various Cosmic Explorer configurations, subdivided in the CE1 and CE2
    stages. CBO and PMO stand for compact-binary or post-merger optimized, i.e.
    focus on low- or high-frequency sensitivity, respectively.}
    \label{fig:psds}
\end{figure}

Figures \ref{fig:psds} and \ref{fig:psd_plots} show the 23 available PSDs
\cite{trade_study_psd} in \gwb. The former is a graphical summary presenting
the hierarchy of these PSDs in terms of detector generation (2G and 3G) and
design concepts (Advanced+ vs Voyager vs ET vs CE and more) while the latter
shows the amplitude spectral densities $\sqrt{S_n(h)}$ as functions of
frequencies. Figure \ref{fig:psd_plots} further presents the observable redshift
ranges for equal-mass binaries with total mass between 1 and 1000 $M_\odot$
(source frame), if observed with SNRs between 12 and 100. Table \ref{tab:tecs}
presents the respective label and key options for these PSDs.

\section{Derivative dictionaries} \label{app:deriv_dicts}

\gwbS obtains the partial derivatives of the WF polarizations and detector responses either
numerically or symbolically using the \textsc{numdifftools} or \textsc{sympy} packages,
respectively. The former calculates the evaluated derivatives directly which is computationally
expensive but applicable to most WFs. \textsc{sympy} on the other hand works with symbolic
expressions that require modifications to the WF source code. In return they enable the analytic
computation of generic derivative expressions which can be stored in the form of \emph{lambdified
functions} for fast evaluation in run time.

\begin{table}[ht]
    \caption{The available detector PSDs are presented together with the respective labels and
    internal keys as used by \gwb. CBO stands for compact-binary optimized and PMO for post-merger
    optimized. The former PSD will be tuned towards increased sensitivity at low frequencies, while
    the latter is geared towards high frequencies. There are two examples illustrating the
    nomenclature of the CE configurations.}
\centering
    \begin{tabular}{c|c|l}
    \hline \hline
    Label & Internal keys & PSD  \\
    \hline \hline
    - & \texttt{aLIGO} & aLIGO \\
    \hline
    + & \texttt{A+,V+,K+} & A+ (adV+/KAGRA+ for V/K location)                 \\
    \hline
    vc & \texttt{Voyager-CBO} & Voyager, CBO                \\
    \hline
    vp & \texttt{Voyager-PMO} & Voyager, PMO                    \\
    \hline
    E & \texttt{ET} & ET-D (same as location) \\
    \hline
    sLo & \texttt{<s>-<L>-<o>} & CE configurations (see below)  \\
    \hline \hline
    \end{tabular}

    \begin{tabular}{l|c|c|c}
    \multicolumn{4}{c}{} \\
    \multicolumn{4}{c}{\multirow{2}{*}{CE configurations - \texttt{<s>-<L>-<o>}}} \\
    \multicolumn{4}{c}{} \\
    \hline \hline
    & stage s & length L & optimizations o \\
    \hline \hline
    Label option & i / a & 1 / 2 / 3 / 4 & c / p \\
    \hline
    Key option & \texttt{CE1} / \texttt{CE2} & \texttt{10} / \texttt{20} / \texttt{30} / \texttt{40} & \texttt{CBO} / \texttt{PMO} \\
    \hline
    PSD setting & CE1 / CE2 & $(10$ / $20$ / $30$ / $40)\,\text{km}$ & CBO / PMO \\
    \hline \hline
    \end{tabular}

    \begin{tabular}{l|l|l}
    \multicolumn{3}{c}{} \\
    \multicolumn{3}{c}{\multirow{2}{*}{CE configurations - Examples}} \\
    \multicolumn{3}{c}{} \\
    \hline \hline
    Label & Internal key & PSD \\
    \hline \hline
    i2p & \texttt{CE1-20-PMO} & CE1, 20 km, PMO \\
    \hline
    a4c & \texttt{CE2-40-CBO} & CE2, 40 km, CBO \\
    \hline \hline
    \end{tabular}
    \label{tab:tecs}
\end{table}

The evaluated partial derivatives are stored in dictionaries which are structured as follows

\begin{lstlisting}
{'del_variable1_hf':<numpy array>,
 'del_variable2_hf':<numpy array>,
  ... }
\end{lstlisting}

\noindent in case of the detector responses and \\ \\

\begin{lstlisting}
{'del_variable1_hfp':<numpy array>,
 'del_variable1_hfc':<numpy array>,
 'del_variable2_hfp':<numpy array>,
 'del_variable2_hfc':<numpy array>,
 ... }
\end{lstlisting}

\noindent for the WF polarizations. These dictionaries have the same structure in case of the
symbolic derivative expressions, but store  lambdified \textsc{sympy} functions instead of
\textsc{numpy} arrays. Further, they contain two extra keys, \texttt{'variables'} and
\texttt{'deriv\_variables'}, that store the input parameters and the differentiation variables.


\section{\textsc{injections} module}\label{app:injections}

The module \textsc{injections} contains functions to randomly sample input parameters appropriate
for GW problems: chirpmass and symmetric mass ratio, binary component spins, sky location and binary
orientation angles, as well as redshifts and luminosity distances. The inputs of the sampler
functions are the same across the board: dictionaries specifying details of the parameter space to
sample, the number of injections \texttt{num\_injs} to sample, and the seed value \texttt{seed} for
the random number generator. In the following we show the usage of these functions and explain what
the respective dictionaries contain.

\paragraph*{Mass sampler:}

The chirpmass and symmetric mass ratio sampling is handled by \texttt{mass\_sampler}:

\begin{lstlisting}
from gwbench import injections

Mcs, etas = injections.mass_sampler(
              mass_dict,num_injs,seed)
\end{lstlisting}

\noindent The \texttt{mass\_dict} contains three definite keys:

\begin{itemize}
        \item \texttt{'dist'}: specifies the mass distribution (values: \texttt{'gaussian'},
        \texttt{'power'}, \texttt{'power\_uniform'}, and \texttt{'uniform'}).
        \item \texttt{'mmin'}: specifies the minimum mass.
        \item \texttt{'mmax'}: specifies the maximum mass.
\end{itemize}

\noindent The Gaussian and power-law distributions need further inputs, namely \texttt{'mean'}
and \texttt{'sigma'} or \texttt{'alpha'}, respectively. The first two specify the mean and standard
deviation for the Gaussian distribution while the \texttt{'alpha'} is the power-law index.

\paragraph*{Spin sampler:}

The binary spin components sampling is handled by \texttt{spin\_sampler}:

\begin{lstlisting}
chi1x, chi1y, chi1z, chi2x, chi2y, chi2z =
    injections.spin_sampler(spin_dict,num_injs,seed)
\end{lstlisting}

\noindent The \texttt{spin\_dict} contains four keys.

\begin{itemize}
        \item \texttt{'geom'}: specifies the geometry (values: \texttt{'cartesian'} and
        \texttt{'spherical'}).
        \item \texttt{'dim'}: specifies spin dimension (values: \texttt{1} and \texttt{3};
        one-dimensional spin sampling disregards \texttt{'geom'} and samples using Cartesian
        geometry; further \texttt{1} returns zeroes for the x- and y-components
        to sample aligned injections).
        \item \texttt{'chi\_lo'}: specifies the minimum spin value (smallest spin value in each
        dimension for \texttt{'cartesian'} geometry or smallest spin magnitude for \texttt{'spherical'}
        geometry).
        \item \texttt{'chi\_hi'}: specifies the maximum spin value (largest spin value in each
        dimension for \texttt{'cartesian'} geometry or largest spin magnitude for \texttt{'spherical'}
        geometry).
\end{itemize}

\paragraph*{Angle sampler:}

The sky location and orientation angles sampling is handled by \texttt{angle\_sampler}:

\begin{lstlisting}
iotas, ras, decs, psis = injections.angle_sampler(
                             num_injs,seed)
\end{lstlisting}

\paragraph*{Redshift and luminosity distance sampler:}

The redshift and luminosity distance sampling is handled by
\texttt{redshift\_lum\_distance\_sampler}:

\begin{lstlisting}
zs, DLs  = injections.redshift_lum_distance_sampler(
            cosmo_dict,num_injs,seed)
\end{lstlisting}

\noindent The \texttt{cosmo\_dict} contains three definite keys

\begin{itemize}
        \item \texttt{'sampler'}: specifies the redshift sampling method (values:
        \texttt{'uniform'}, \texttt{'uniform\_comoving\_volume\_inversion'} and \texttt{'uniform\_comoving\_volume\_rejection'}; 
        the inversion method is faster, while the rejection method is more accurate, especially for very jagged redshift distributions).
        \item \texttt{'zmin'}: specifies the minimum redshift.
        \item \texttt{'zmax'}: specifies the maximum redshift.
\end{itemize}

By default the code samples redshifts from the redshift distribution
\texttt{astropy.cosmology.Planck18} provided in the \textsc{astropy} package. The user can change
this behavior by adding three further keys to the \texttt{cosmo\_dict} whose values are then passed
to \texttt{astropy.cosmology.LambdaCDM}:

\begin{itemize}
        \item \texttt{'H0'}: specifies Hubble constant.
        \item \texttt{'Om0'}: specifies the mass density $\Omega_{m,0}$ at the current time.
        \item \texttt{'Ode0'}: specifies dark energy density $\Omega_{\Lambda,0}$ at the current time.
\end{itemize}

\paragraph*{Injection parameter sampler:}

\textsc{injections} further contains a function that combines the four other samplers into one
function call: \texttt{injections\_CBC\_params\_redshift}. Hence it depends on the same input as the
previous four methods:

\begin{lstlisting}
data  = injections.injections_CBC_params_redshift(
        cosmo_dict,mass_dict,spin_dict,redshifted,
        num_injs,seed,file_path)
\end{lstlisting}

\noindent The two additional inputs, \texttt{redshifted} and \texttt{file\_path}, control whether
the sampled masses are already redshifted (\texttt{1}) according to the sampled redshifts or not
(\texttt{0}) as well as if the sampled data should be stored in a file at the specified path
(default is \texttt{None}), respectively.


\bibliographystyle{JHEP}
\bibliography{refs}

\providecommand{\href}[2]{#2}\begingroup\raggedright\begin{thebibliography}{10}

\bibitem{GW150914}
{\scshape LIGO Scientific, Virgo} collaboration, \emph{{Observation of
  Gravitational Waves from a Binary Black Hole Merger}},
  \href{https://doi.org/10.1103/PhysRevLett.116.061102}{\emph{Phys. Rev. Lett.}
  {\bfseries 116} (2016) 061102}
  [\href{https://arxiv.org/abs/1602.03837}{{\ttfamily 1602.03837}}].

\bibitem{GW170817}
{\scshape LIGO Scientific, Virgo} collaboration, \emph{{GW170817: Observation
  of Gravitational Waves from a Binary Neutron Star Inspiral}},
  \href{https://doi.org/10.1103/PhysRevLett.119.161101}{\emph{Phys. Rev. Lett.}
  {\bfseries 119} (2017) 161101}
  [\href{https://arxiv.org/abs/1710.05832}{{\ttfamily 1710.05832}}].

\bibitem{GWTC1}
{\scshape LIGO Scientific, Virgo} collaboration, \emph{{GWTC-1: A
  Gravitational-Wave Transient Catalog of Compact Binary Mergers Observed by
  LIGO and Virgo during the First and Second Observing Runs}},
  \href{https://doi.org/10.1103/PhysRevX.9.031040}{\emph{Phys. Rev. X}
  {\bfseries 9} (2019) 031040}
  [\href{https://arxiv.org/abs/1811.12907}{{\ttfamily 1811.12907}}].

\bibitem{London:2017bcn}
L.~London, S.~Khan, E.~Fauchon-Jones, C.~Garc\'\i{}a, M.~Hannam, S.~Husa
  et~al., \emph{{First higher-multipole model of gravitational waves from
  spinning and coalescing black-hole binaries}},
  \href{https://doi.org/10.1103/PhysRevLett.120.161102}{\emph{Phys. Rev. Lett.}
  {\bfseries 120} (2018) 161102}
  [\href{https://arxiv.org/abs/1708.00404}{{\ttfamily 1708.00404}}].

\bibitem{Cotesta:2018fcv}
R.~Cotesta, A.~Buonanno, A.~Boh\'e, A.~Taracchini, I.~Hinder and S.~Ossokine,
  \emph{{Enriching the Symphony of Gravitational Waves from Binary Black Holes
  by Tuning Higher Harmonics}},
  \href{https://doi.org/10.1103/PhysRevD.98.084028}{\emph{Phys. Rev. D}
  {\bfseries 98} (2018) 084028}
  [\href{https://arxiv.org/abs/1803.10701}{{\ttfamily 1803.10701}}].

\bibitem{aLIGO}
{\scshape LIGO Scientific} collaboration, \emph{{Advanced LIGO}},
  \href{https://doi.org/10.1088/0264-9381/32/7/074001}{\emph{Class. Quant.
  Grav.} {\bfseries 32} (2015) 074001}
  [\href{https://arxiv.org/abs/1411.4547}{{\ttfamily 1411.4547}}].

\bibitem{Virgo}
{\scshape VIRGO} collaboration, \emph{{Advanced Virgo: a second-generation
  interferometric gravitational wave detector}},
  \href{https://doi.org/10.1088/0264-9381/32/2/024001}{\emph{Class. Quant.
  Grav.} {\bfseries 32} (2015) 024001}
  [\href{https://arxiv.org/abs/1408.3978}{{\ttfamily 1408.3978}}].

\bibitem{KAGRA}
{\scshape KAGRA} collaboration, \emph{{Interferometer design of the KAGRA
  gravitational wave detector}},
  \href{https://doi.org/10.1103/PhysRevD.88.043007}{\emph{Phys. Rev. D}
  {\bfseries 88} (2013) 043007}
  [\href{https://arxiv.org/abs/1306.6747}{{\ttfamily 1306.6747}}].

\bibitem{ligo_india}
C.~for Indian Initiative in Gravitational-wave Observations~(IndIGO),
  \emph{Ligo-india, proposal},  (2011),
  \href{https://dcc.ligo.org/LIGO-M1100296/public}{https://dcc.ligo.org/LIGO-M1100296/public}.

\bibitem{aPlusWP}
LSC, \emph{Instrument science white paper},  (2016),
  \href{https://dcc.ligo.org/LIGO-T1600119/public}{https://dcc.ligo.org/LIGO-T1600119/public}.

\bibitem{Aasi:2013wya}
{\scshape KAGRA, LIGO Scientific, VIRGO} collaboration, \emph{{Prospects for
  Observing and Localizing Gravitational-Wave Transients with Advanced LIGO,
  Advanced Virgo and KAGRA}},
  \href{https://doi.org/10.1007/s41114-018-0012-9}{\emph{Living Rev. Rel.}
  {\bfseries 21} (2018) 3} [\href{https://arxiv.org/abs/1304.0670}{{\ttfamily
  1304.0670}}].

\bibitem{LISA_science_case}
{\scshape eLISA} collaboration, \emph{{The Gravitational Universe}},
  [\href{https://arxiv.org/abs/1305.5720}{{\ttfamily 1305.5720}}].

\bibitem{ET_science_case}
M.~Maggiore et~al., \emph{{Science Case for the Einstein Telescope}},
  \href{https://doi.org/10.1088/1475-7516/2020/03/050}{\emph{JCAP} {\bfseries
  03} (2020) 050} [\href{https://arxiv.org/abs/1912.02622}{{\ttfamily
  1912.02622}}].

\bibitem{Voyager}
{\scshape LIGO} collaboration, \emph{{A Cryogenic Silicon Interferometer for
  Gravitational-wave Detection}},
  \href{https://doi.org/10.1088/1361-6382/ab9143}{\emph{Class. Quant. Grav.}
  {\bfseries 37} (2020) 165003}
  [\href{https://arxiv.org/abs/2001.11173}{{\ttfamily 2001.11173}}].

\bibitem{ET_1}
M.~Punturo et~al., \emph{{The Einstein Telescope: A third-generation
  gravitational wave observatory}},
  \href{https://doi.org/10.1088/0264-9381/27/19/194002}{\emph{Class. Quant.
  Grav.} {\bfseries 27} (2010) 194002}.

\bibitem{ET_2}
M.~Punturo et~al., \emph{{The third generation of gravitational wave
  observatories and their science reach}},
  \href{https://doi.org/10.1088/0264-9381/27/8/084007}{\emph{Class. Quant.
  Grav.} {\bfseries 27} (2010) 084007}.

\bibitem{CE_proposal}
D.~Reitze et~al., \emph{{Cosmic Explorer: The U.S. Contribution to
  Gravitational-Wave Astronomy beyond LIGO}}, {\emph{Bull. Am. Astron. Soc.}
  {\bfseries 51} (2019) 035}
  [\href{https://arxiv.org/abs/1907.04833}{{\ttfamily 1907.04833}}].

\bibitem{Cutler:1994ys}
C.~Cutler and E.~E. Flanagan, \emph{{Gravitational waves from merging compact
  binaries: How accurately can one extract the binary's parameters from the
  inspiral wave form?}},
  \href{https://doi.org/10.1103/PhysRevD.49.2658}{\emph{Phys. Rev. D}
  {\bfseries 49} (1994) 2658}
  [\href{https://arxiv.org/abs/gr-qc/9402014}{{\ttfamily gr-qc/9402014}}].

\bibitem{Poisson:1995ef}
E.~Poisson and C.~M. Will, \emph{{Gravitational waves from inspiraling compact
  binaries: Parameter estimation using second postNewtonian wave forms}},
  \href{https://doi.org/10.1103/PhysRevD.52.848}{\emph{Phys. Rev. D} {\bfseries
  52} (1995) 848} [\href{https://arxiv.org/abs/gr-qc/9502040}{{\ttfamily
  gr-qc/9502040}}].

\bibitem{Vallisneri:2007ev}
M.~Vallisneri, \emph{{Use and abuse of the Fisher information matrix in the
  assessment of gravitational-wave parameter-estimation prospects}},
  \href{https://doi.org/10.1103/PhysRevD.77.042001}{\emph{Phys. Rev. D}
  {\bfseries 77} (2008) 042001}
  [\href{https://arxiv.org/abs/gr-qc/0703086}{{\ttfamily gr-qc/0703086}}].

\bibitem{Chua:2019wwt}
A.~J.~K. Chua and M.~Vallisneri, \emph{{Learning Bayesian posteriors with
  neural networks for gravitational-wave inference}},
  \href{https://doi.org/10.1103/PhysRevLett.124.041102}{\emph{Phys. Rev. Lett.}
  {\bfseries 124} (2020) 041102}
  [\href{https://arxiv.org/abs/1909.05966}{{\ttfamily 1909.05966}}].

\bibitem{Gabbard:2019rde}
H.~Gabbard, C.~Messenger, I.~S. Heng, F.~Tonolini and R.~Murray-Smith,
  \emph{{Bayesian parameter estimation using conditional variational
  autoencoders for gravitational-wave astronomy}},
  [\href{https://arxiv.org/abs/1909.06296}{{\ttfamily 1909.06296}}].

\bibitem{Green:2020dnx}
S.~R. Green and J.~Gair, \emph{{Complete parameter inference for GW150914 using
  deep learning}}, \href{https://doi.org/10.1088/2632-2153/abfaed}{\emph{Mach.
  Learn. Sci. Tech.} {\bfseries 2} (2021) 03LT01}
  [\href{https://arxiv.org/abs/2008.03312}{{\ttfamily 2008.03312}}].

\bibitem{Green:2020hst}
S.~R. Green, C.~Simpson and J.~Gair, \emph{{Gravitational-wave parameter
  estimation with autoregressive neural network flows}},
  \href{https://doi.org/10.1103/PhysRevD.102.104057}{\emph{Phys. Rev. D}
  {\bfseries 102} (2020) 104057}
  [\href{https://arxiv.org/abs/2002.07656}{{\ttfamily 2002.07656}}].

\bibitem{lalsuite}
L.~S. Collaboration et~al., \emph{Lsc algorithm library suite},  (2020),
  \href{https://doi.org/10.7935/GT1W-FZ16}{https://doi.org/10.7935/GT1W-FZ16}.

\bibitem{wette2020}
K.~Wette, \emph{{SWIGLAL}: Python and octave interfaces to the lalsuite
  gravitational-wave data analysis libraries}, {\emph{Accepted in SoftwareX}
  (2020) }.

\bibitem{Sathyaprakash:2019yqt}
B.~Sathyaprakash et~al., \emph{{Extreme Gravity and Fundamental Physics}},
  [\href{https://arxiv.org/abs/1903.09221}{{\ttfamily 1903.09221}}].

\bibitem{Kalogera:2019sui}
V.~Kalogera et~al., \emph{{Deeper, Wider, Sharper: Next-Generation Ground-Based
  Gravitational-Wave Observations of Binary Black Holes}},
  [\href{https://arxiv.org/abs/1903.09220}{{\ttfamily 1903.09220}}].

\bibitem{Sathyaprakash:2019rom}
B.~Sathyaprakash et~al., \emph{{Multimessenger Universe with Gravitational
  Waves from Binaries}},  [\href{https://arxiv.org/abs/1903.09277}{{\ttfamily
  1903.09277}}].

\bibitem{Sathyaprakash:2019nnu}
B.~Sathyaprakash et~al., \emph{{Cosmology and the Early Universe}},
  [\href{https://arxiv.org/abs/1903.09260}{{\ttfamily 1903.09260}}].

\bibitem{Barack:2018yly}
L.~Barack et~al., \emph{{Black holes, gravitational waves and fundamental
  physics: a roadmap}},
  \href{https://doi.org/10.1088/1361-6382/ab0587}{\emph{Class. Quant. Grav.}
  {\bfseries 36} (2019) 143001}
  [\href{https://arxiv.org/abs/1806.05195}{{\ttfamily 1806.05195}}].

\bibitem{LRR_Sathyaprakash_2009}
B.~Sathyaprakash and B.~Schutz, \emph{{Physics, Astrophysics and Cosmology with
  Gravitational Waves}},
  \href{https://doi.org/10.12942/lrr-2009-2}{\emph{Living Rev. Rel.} {\bfseries
  12} (2009) 2} [\href{https://arxiv.org/abs/0903.0338}{{\ttfamily
  0903.0338}}].

\bibitem{Wen:2010cr}
L.~Wen and Y.~Chen, \emph{{Geometrical Expression for the Angular Resolution of
  a Network of Gravitational-Wave Detectors}},
  \href{https://doi.org/10.1103/PhysRevD.81.082001}{\emph{Phys. Rev. D}
  {\bfseries 81} (2010) 082001}
  [\href{https://arxiv.org/abs/1003.2504}{{\ttfamily 1003.2504}}].

\bibitem{Zhao:2017cbb}
W.~Zhao and L.~Wen, \emph{{Localization accuracy of compact binary coalescences
  detected by the third-generation gravitational-wave detectors and implication
  for cosmology}},
  \href{https://doi.org/10.1103/PhysRevD.97.064031}{\emph{Phys. Rev. D}
  {\bfseries 97} (2018) 064031}
  [\href{https://arxiv.org/abs/1710.05325}{{\ttfamily 1710.05325}}].

\bibitem{Rodriguez:2011aa}
C.~L. Rodriguez, I.~Mandel and J.~R. Gair, \emph{{Verifying the no-hair
  property of massive compact objects with intermediate-mass-ratio inspirals in
  advanced gravitational-wave detectors}},
  \href{https://doi.org/10.1103/PhysRevD.85.062002}{\emph{Phys. Rev. D}
  {\bfseries 85} (2012) 062002}
  [\href{https://arxiv.org/abs/1112.1404}{{\ttfamily 1112.1404}}].

\bibitem{1991PhRvD..44.3819S}
B.~S. {Sathyaprakash} and S.~V. {Dhurandhar}, \emph{{Choice of filters for the
  detection of gravitational waves from coalescing binaries}},
  \href{https://doi.org/10.1103/PhysRevD.44.3819}{\emph{\prd} {\bfseries 44}
  (1991) 3819}.

\bibitem{Poisson:1997ha}
E.~Poisson, \emph{{Gravitational waves from inspiraling compact binaries: The
  Quadrupole moment term}},
  \href{https://doi.org/10.1103/PhysRevD.57.5287}{\emph{Phys. Rev. D}
  {\bfseries 57} (1998) 5287}
  [\href{https://arxiv.org/abs/gr-qc/9709032}{{\ttfamily gr-qc/9709032}}].

\bibitem{Mikoczi:2005dn}
B.~Mikoczi, M.~Vasuth and L.~A. Gergely, \emph{{Self-interaction spin effects
  in inspiralling compact binaries}},
  \href{https://doi.org/10.1103/PhysRevD.71.124043}{\emph{Phys. Rev. D}
  {\bfseries 71} (2005) 124043}
  [\href{https://arxiv.org/abs/astro-ph/0504538}{{\ttfamily
  astro-ph/0504538}}].

\bibitem{Arun:2008kb}
K.~Arun, A.~Buonanno, G.~Faye and E.~Ochsner, \emph{{Higher-order spin effects
  in the amplitude and phase of gravitational waveforms emitted by inspiraling
  compact binaries: Ready-to-use gravitational waveforms}},
  \href{https://doi.org/10.1103/PhysRevD.79.104023}{\emph{Phys. Rev. D}
  {\bfseries 79} (2009) 104023}
  [\href{https://arxiv.org/abs/0810.5336}{{\ttfamily 0810.5336}}].

\bibitem{Buonanno:2009zt}
A.~Buonanno, B.~Iyer, E.~Ochsner, Y.~Pan and B.~Sathyaprakash,
  \emph{{Comparison of post-Newtonian templates for compact binary inspiral
  signals in gravitational-wave detectors}},
  \href{https://doi.org/10.1103/PhysRevD.80.084043}{\emph{Phys. Rev. D}
  {\bfseries 80} (2009) 084043}
  [\href{https://arxiv.org/abs/0907.0700}{{\ttfamily 0907.0700}}].

\bibitem{Bohe:2013cla}
A.~Boh\'e, S.~Marsat and L.~Blanchet, \emph{{Next-to-next-to-leading order
  spin\textendash{}orbit effects in the gravitational wave flux and orbital
  phasing of compact binaries}},
  \href{https://doi.org/10.1088/0264-9381/30/13/135009}{\emph{Class. Quant.
  Grav.} {\bfseries 30} (2013) 135009}
  [\href{https://arxiv.org/abs/1303.7412}{{\ttfamily 1303.7412}}].

\bibitem{Bohe:2015ana}
A.~Boh\'e, G.~Faye, S.~Marsat and E.~K. Porter, \emph{{Quadratic-in-spin
  effects in the orbital dynamics and gravitational-wave energy flux of compact
  binaries at the 3PN order}},
  \href{https://doi.org/10.1088/0264-9381/32/19/195010}{\emph{Class. Quant.
  Grav.} {\bfseries 32} (2015) 195010}
  [\href{https://arxiv.org/abs/1501.01529}{{\ttfamily 1501.01529}}].

\bibitem{Mishra:2016whh}
C.~K. Mishra, A.~Kela, K.~Arun and G.~Faye, \emph{{Ready-to-use post-Newtonian
  gravitational waveforms for binary black holes with nonprecessing spins: An
  update}}, \href{https://doi.org/10.1103/PhysRevD.93.084054}{\emph{Phys. Rev.
  D} {\bfseries 93} (2016) 084054}
  [\href{https://arxiv.org/abs/1601.05588}{{\ttfamily 1601.05588}}].

\bibitem{astropy:2013}
{\scshape Astropy} collaboration, \emph{{Astropy: A Community Python Package
  for Astronomy}},
  \href{https://doi.org/10.1051/0004-6361/201322068}{\emph{Astron. Astrophys.}
  {\bfseries 558} (2013) A33}
  [\href{https://arxiv.org/abs/1307.6212}{{\ttfamily 1307.6212}}].

\bibitem{astropy:2018}
{\scshape Astropy} collaboration, \emph{{The Astropy Project: Building an
  Open-science Project and Status of the v2.0 Core Package}},
  \href{https://doi.org/10.3847/1538-3881/aabc4f}{\emph{Astron. J.} {\bfseries
  156} (2018) 123} [\href{https://arxiv.org/abs/1801.02634}{{\ttfamily
  1801.02634}}].

\bibitem{numdifftools}
P.~A. Brodtkorb and J.~D’Errico, \emph{Documentation: numdifftools},  (2019),
  \href{https://numdifftools.readthedocs.io/en/latest/index.html}{https://numdifftools.readthedocs.io/en/latest/index.html}.

\bibitem{GW190521}
{\scshape LIGO Scientific, Virgo} collaboration, \emph{{GW190521: A Binary
  Black Hole Merger with a Total Mass of $150 ~ M_{\odot}$}},
  \href{https://doi.org/10.1103/PhysRevLett.125.101102}{\emph{Phys. Rev. Lett.}
  {\bfseries 125} (2020) 101102}
  [\href{https://arxiv.org/abs/2009.01075}{{\ttfamily 2009.01075}}].

\bibitem{GW190814}
{\scshape LIGO Scientific, Virgo} collaboration, \emph{{GW190814: Gravitational
  Waves from the Coalescence of a 23 Solar Mass Black Hole with a 2.6 Solar
  Mass Compact Object}},
  \href{https://doi.org/10.3847/2041-8213/ab960f}{\emph{Astrophys. J. Lett.}
  {\bfseries 896} (2020) L44}
  [\href{https://arxiv.org/abs/2006.12611}{{\ttfamily 2006.12611}}].

\bibitem{Lackey:2014fwa}
B.~D. Lackey and L.~Wade, \emph{{Reconstructing the neutron-star equation of
  state with gravitational-wave detectors from a realistic population of
  inspiralling binary neutron stars}},
  \href{https://doi.org/10.1103/PhysRevD.91.043002}{\emph{Phys. Rev. D}
  {\bfseries 91} (2015) 043002}
  [\href{https://arxiv.org/abs/1410.8866}{{\ttfamily 1410.8866}}].

\bibitem{Wade:2014vqa}
L.~Wade, J.~D. Creighton, E.~Ochsner, B.~D. Lackey, B.~F. Farr, T.~B.
  Littenberg et~al., \emph{{Systematic and statistical errors in a bayesian
  approach to the estimation of the neutron-star equation of state using
  advanced gravitational wave detectors}},
  \href{https://doi.org/10.1103/PhysRevD.89.103012}{\emph{Phys. Rev. D}
  {\bfseries 89} (2014) 103012}
  [\href{https://arxiv.org/abs/1402.5156}{{\ttfamily 1402.5156}}].

\bibitem{GWTC1_PSDs}
LSC, \emph{Power spectral densities (psd) release for gwtc-1},  (2019),
  \href{https://dcc.ligo.org/LIGO-P1900011/public}{https://dcc.ligo.org/LIGO-P1900011/public}.

\bibitem{Husa:2015iqa}
S.~Husa, S.~Khan, M.~Hannam, M.~P\"urrer, F.~Ohme, X.~Jim\'enez~Forteza et~al.,
  \emph{{Frequency-domain gravitational waves from nonprecessing black-hole
  binaries. I. New numerical waveforms and anatomy of the signal}},
  \href{https://doi.org/10.1103/PhysRevD.93.044006}{\emph{Phys. Rev. D}
  {\bfseries 93} (2016) 044006}
  [\href{https://arxiv.org/abs/1508.07250}{{\ttfamily 1508.07250}}].

\bibitem{Khan:2015jqa}
S.~Khan, S.~Husa, M.~Hannam, F.~Ohme, M.~P\"urrer, X.~Jim\'enez~Forteza et~al.,
  \emph{{Frequency-domain gravitational waves from nonprecessing black-hole
  binaries. II. A phenomenological model for the advanced detector era}},
  \href{https://doi.org/10.1103/PhysRevD.93.044007}{\emph{Phys. Rev. D}
  {\bfseries 93} (2016) 044007}
  [\href{https://arxiv.org/abs/1508.07253}{{\ttfamily 1508.07253}}].

\bibitem{Borhanian:2020vyr}
S.~Borhanian, A.~Dhani, A.~Gupta, K.~G. Arun and B.~S. Sathyaprakash,
  \emph{{Dark Sirens to Resolve the Hubble\textendash{}Lema\^\i{}tre Tension}},
  \href{https://doi.org/10.3847/2041-8213/abcaf5}{\emph{Astrophys. J. Lett.}
  {\bfseries 905} (2020) L28}
  [\href{https://arxiv.org/abs/2007.02883}{{\ttfamily 2007.02883}}].

\bibitem{Gupta:2020lxa}
A.~Gupta, S.~Datta, S.~Kastha, S.~Borhanian, K.~Arun and B.~Sathyaprakash,
  \emph{{Multiparameter tests of general relativity using multiband
  gravitational-wave observations}},
  \href{https://doi.org/10.1103/PhysRevLett.125.201101}{\emph{Phys. Rev. Lett.}
  {\bfseries 125} (2020) 201101}
  [\href{https://arxiv.org/abs/2005.09607}{{\ttfamily 2005.09607}}].

\bibitem{Arun:2006hn}
K.~Arun, B.~R. Iyer, M.~Qusailah and B.~Sathyaprakash, \emph{{Probing the
  non-linear structure of general relativity with black hole binaries}},
  \href{https://doi.org/10.1103/PhysRevD.74.024006}{\emph{Phys. Rev. D}
  {\bfseries 74} (2006) 024006}
  [\href{https://arxiv.org/abs/gr-qc/0604067}{{\ttfamily gr-qc/0604067}}].

\bibitem{gstlal_live}
C.~Messick et~al., \emph{{Analysis Framework for the Prompt Discovery of
  Compact Binary Mergers in Gravitational-wave Data}},
  \href{https://doi.org/10.1103/PhysRevD.95.042001}{\emph{Phys. Rev. D}
  {\bfseries 95} (2017) 042001}
  [\href{https://arxiv.org/abs/1604.04324}{{\ttfamily 1604.04324}}].

\bibitem{pycbc_live}
A.~H. Nitz, T.~Dal~Canton, D.~Davis and S.~Reyes, \emph{{Rapid detection of
  gravitational waves from compact binary mergers with PyCBC Live}},
  \href{https://doi.org/10.1103/PhysRevD.98.024050}{\emph{Phys. Rev. D}
  {\bfseries 98} (2018) 024050}
  [\href{https://arxiv.org/abs/1805.11174}{{\ttfamily 1805.11174}}].

\bibitem{Moore:2019pke}
C.~J. Moore, D.~Gerosa and A.~Klein, \emph{{Are stellar-mass black-hole
  binaries too quiet for LISA?}},
  \href{https://doi.org/10.1093/mnrasl/slz104}{\emph{Mon. Not. Roy. Astron.
  Soc.} {\bfseries 488} (2019) L94}
  [\href{https://arxiv.org/abs/1905.11998}{{\ttfamily 1905.11998}}].

\bibitem{Owen:1995tm}
B.~J. Owen, \emph{{Search templates for gravitational waves from inspiraling
  binaries: Choice of template spacing}},
  \href{https://doi.org/10.1103/PhysRevD.53.6749}{\emph{Phys. Rev. D}
  {\bfseries 53} (1996) 6749}
  [\href{https://arxiv.org/abs/gr-qc/9511032}{{\ttfamily gr-qc/9511032}}].

\bibitem{Owen:1998dk}
B.~J. Owen and B.~Sathyaprakash, \emph{{Matched filtering of gravitational
  waves from inspiraling compact binaries: Computational cost and template
  placement}}, \href{https://doi.org/10.1103/PhysRevD.60.022002}{\emph{Phys.
  Rev. D} {\bfseries 60} (1999) 022002}
  [\href{https://arxiv.org/abs/gr-qc/9808076}{{\ttfamily gr-qc/9808076}}].

\bibitem{Schutz:2011tw}
B.~F. Schutz, \emph{{Networks of gravitational wave detectors and three figures
  of merit}},
  \href{https://doi.org/10.1088/0264-9381/28/12/125023}{\emph{Class. Quant.
  Grav.} {\bfseries 28} (2011) 125023}
  [\href{https://arxiv.org/abs/1102.5421}{{\ttfamily 1102.5421}}].

\bibitem{trade_study_psd}
K.~Kuns, V.~Srivastava, E.~Hall, M.~Evans and S.~Ballmer (2020),
  \href{https://dcc.cosmicexplorer.org/CE-T2000007}{https://dcc.cosmicexplorer.org/CE-T2000007}.

\end{thebibliography}\endgroup
\end{document}